\documentclass[journal]{IEEEtran}

\ifCLASSINFOpdf
  \usepackage[pdftex]{graphicx}
\else
\fi

\usepackage{amsmath}
\DeclareMathOperator*{\argmax}{arg\,max}

\usepackage{amsfonts}

\usepackage[nolist,nohyperlinks]{acronym}

\usepackage[hyphens]{url}
\usepackage[center]{caption}
\usepackage{multirow}
\usepackage{multicol}
\usepackage{relsize}

\usepackage[ruled,vlined]{algorithm2e}
\usepackage{amssymb}
\usepackage[table]{xcolor}
\usepackage{bm}

\usepackage{dblfloatfix}

\begin{document}
\title{Uncertainty Quantification in Machine Learning for Joint Speaker Diarization and Identification}

\author{Simon~W.~McKnight,~\IEEEmembership{Student~Member,~IEEE,}
        Aidan~O.~T.~Hogg,~\IEEEmembership{Student~Member,~IEEE,}
        Vincent~W.~Neo,~\IEEEmembership{Student~Member,~IEEE,}
        and~Patrick~A.~Naylor,~\IEEEmembership{Fellow,~IEEE}% <-this % stops a space
\thanks{The authors are in the Electrical and Electronic Engineering Department, Imperial College London, Exhibition Road, South Kensington, London SW7 2BX, UK, e-mails: \{s.mcknight18, a.hogg, vincent.neo09, p.naylor\}@imperial.ac.uk.}% <-this % stops a space
\thanks{This manuscript is first received 5 May 2023.}}

\begin{acronym}
%% List of SAP acronyms. Some non-obvious points to note are:
%% (1) Keep in alphabetical order to print correctly.
%%     The perl script "tidy_acronyms.pl" will sort the acronyms.
%% (2) Put each complete acronym definition in a single line
%% (3) No non-standard characters. Accents should use the LaTeX escapes.
%%     The perl script "tidy_acronyms.pl" will replace accents with the correct form.
%% (4) Use \acs{...} to include another acronym in the definition
%%     e.g. \acro{SSNR}{Segmental \acs{SNR}}
%% (5) Text additional to the definition should be enclosed thus: \acroextra{ }
%%     e.g. \acro{AAC}{Advanced Audio Coding\acroextra{. A lossy codec used for digital audio.}}
%% (6) Multiple meanings of an acronym are distinguished thus:
%%     e.g. \acro{SNR1}[SNR]{Society for Nautical Research}
%% (7) You must also use this form if the the acronym involves mathematical symbols:
%%     e.g. \acro{Leq}[L$_{\textrm{eq}}$]{Equivalent Continuous Sound Level}
%% (8) Non-standard plurals can be defined as:
%%     e.g. \acro{MP}{Members of Parliament}\acrodefplural{MP}[MPs]{Members of Parliament}
%% (9) Different indefinite articles can be specified for long and short forms
%%     e.g. \acro{FFT}{Fast Fourier Transform}\acroindefinite{FFT}{an}{a}
%%
\acro{3GPP}{3rd Generation Partnership Project}
\acro{a-SLAM}[aSLAM]{Acoustic \aclu{SLAM}}
\acro{aSLAM}{Acoustic \aclu{SLAM}}
\acro{A-SNR}{A-weighted \acs{SNR}}
\acro{AAC}{Advanced Audio Coding\acroextra{. A lossy codec used for digital audio.}}
\acro{AAD}{Auditory Attention Detection}
\acro{AAS}{American Auditory Soc.}
\acro{AASP}{Audio and Acoustic Signal Processing}
\acro{ABC}{Analytical with or without Bias Compensation}
\acro{ABR}{Auditory-Brainstem Response}
\acro{ACAWD}{Archivable Core Actual-Word Database}
\acro{ACB}{Adaptive Codebook}
\acro{ACC}{Accuracy}
\acro{ACE}{Acoustic Characterization of Environments\acroextra{. A noisy reverberant speech corpus and IEEE challenge run by the SAP group at Imperial College}}
\acro{ACELP}{Algebraic Code-Excited Linear Prediction}
\acro{ACF}{Autocorrelation Function}
\acro{ACK}{Acknowledgement}
\acro{ACL}{Access Control List}
\acro{ACR}{Absolute Category Rating}
\acro{AD}{Audio Diarization}
\acro{ADC}{Analogue-to-Digital Converter}
\acro{ADM}{Adaptive Differential Microphone}
\acro{ADPCM}{Adaptive Differential Pulse Code Modulation}
\acro{ADSL}{Asymmetric Digital Subscriber Line}
\acro{AE}{Almost Everywhere}
\acro{AES}{Audio Engineering Society}
\acro{AES2}[AES]{Advanced Encryption Standard}
\acro{AGC}{Automatic Gain Control}
\acro{AH}{Amplitude Histogram}
\acro{AHC}{agglomerative hierarchical clustering}
%\acro{AI}{Articulation Index}
%\acro{AI2}[AI]{Artificial Intelligence}
%\acro{AI3}[AI]{Audio Inpainting}
\acro{AI}{artificial intelligence}
\acro{aIB}{agglomerative information bottleneck}
\acro{AIC}{Akaike Information Criterion}
\acro{AIFF}{Audio Interchange File Format}
\acro{AIR}{Acoustic Impulse Response}
\acro{AIR2}[AIR]{Aachen Impulse Response}
\acro{AIRD}{Aachen Impulse Response Database}
\acro{ALC}{Automatic Level Control}
\acro{ALCons}{Articulation Loss of Consonants}
\acro{AM}{amplitude modulation}
\acro{AMDF}{Average Magnitude Difference Function\acroextra{. A function with similar properties to the cross- or autocorrelation but that requires no multiplication to evaluate.}}
\acro{AMI}{Augmented Multi-party Interaction}
\acro{AMR}{Adaptive Multi-Rate}
\acro{AMR-NB}{Adaptive Multi-Rate Narrow Band}
\acro{AMR-WB}{Adaptive Multi-Rate Wide Band}
% \acro{AMS}{Amplitude Modulation Spectrogram}
\acro{AMS}{analysis-modification-synthesis}
\acro{ANC}{Adaptive Noise Canceller}
\acro{ANOVA}{analysis of variance}
\acro{ANS}{Autocorrelation-Based Noise Subtraction}
\acro{ANSI}{American National Standards Institute}
\acro{ANU}{Australian National University}
\acro{APLAWD}{Archivable Priority List Actual-Word Database}
\acro{API}{application programming interface}
\acro{AR}{Autoregressive}
%% "ARD" changed by dmb to avoid errors
%% previously: \acro{ARD}{Arbeitsgemeinschaft der ffentlich-rechtlichen Rundfunkanstalten der Bundesrepublik Deutschland\acroextra{. 'Consortium' ("Working group") Ð of the public-law broadcasting institutions Ð of the Federal Republic of GermanyÕ}}
%\acro{ARD}{Arbeitsgemeinschaft der \"{o}ffentlich-rechtlichen Rundfunkanstalten der Bundesrepublik Deutschland\acroextra{. `Consortium (``Working group'') of the public-law broadcasting institutions of the Federal Republic of Germany'}}
\acro{ARD}{automatic relevance determination}
\acro{ARMA}{Autoregressive Moving Average}
\acro{AS}{Audio Segmentation}
\acro{AS2}[AS]{Almost Surely}
\acro{ASA}{Acoustic Scene Analysis}
\acro{ASIO}{Audio Stream Input/Output\acroextra{. A computer soundcard protocol with low latency developed by Steinberg.}}
\acro{ASK}{Amplitude Shift Keying}
\acro{ASLP}{Audio, Speech, and Language Processing}
\acro{ASM}{Acoustic Scene Mapping}
\acro{ASR}{automatic speech recognition}
\acro{ASS}{Approximate Spectrum Substitution}
\acro{AST}{Acoustic Source Tracking}
\acro{AST2}[AST]{Asymmetric Sampling in Time}
\acro{ATF}{Acoustic Transfer Function\acroextra{. The Fourier Transform of the \acs{RIR}.}}
\acro{ATLM}{Acoustic Tokenization and Language Modelling}
\acro{ATR}{Advanced Telecommunications Research Institute International\acroextra{, Kyoto, Japan}}
\acro{AUC}{Area under the curve}
\acro{AURORA}{Aurora Experimental Framework for the Evaluation of the Performance of Speech Recognition Systems under Noisy Conditions}
\acro{AUV}{Autonomous Underwater Vehicle}
\acro{AV}{Audio-Visual}
\acro{AWGN}{additive white Gaussian noise}
\acro{AWS}{Approximate Waveform Substitution}
\acro{BASIE}{Bayesian Adaptive Speech Intelligibility Estimation}
% \acro{BCE}{Blind Channel Estimation}
\acro{BCE}{binary cross-entropy}
\acro{BCL}{Bekesy Comfortable Loudness}
\acro{BCR}{Block-Coordinate Relaxation}
\acro{BEM}{Boundary Element Method}
\acro{BER}{Bit Error Rate}
\acro{BIBO}{Bounded-Input Bounded-Output}
\acro{BIC}{Bayesian information criterion}
\acro{Bk}{Berksons\acroextra{. A unit for measuring intelligibility.}}
\acro{BK}[B\&K]{Br{\"u}el and Kj{\ae}r}
\acro{BiLSTM}{Bidirectional \acs{LSTM}}
\acro{BM}{Blocking Matrix}
\acro{BNN}{Bayesian neural network}
\acro{BNNs}{Bayesian neural networks}
\acro{BO-SCPHD}{Bearing-only \acs{SC-PHD}}
\acro{BO-SLAM}{Bearing-only \acs{SLAM}}
\acro{BOT}{Bearing-only tracking}
\acro{BP}{Basis Pursuit}
\acro{BPCC}{Basis Pursuit with Clipping Constraints}
\acro{BPDN}{Basis Pursuit Denoising}
\acro{BPM}{Beats Per Minute}
\acro{BPSK}{Binary Phase Shift Keying}
\acro{BR}{Barrodale and Roberts' (algorithm)}
\acro{BRI}{Basic Rate Index}
\acro{BSD}{Bark Spectral Distortion}
\acro{BSI}{Blind System Identification}
\acro{BSIM}{Binaural Speech Intelligbility Model}
\acro{BSS}{Blind Source Separation}
\acro{BSTOI}{Binaural \acs{STOI}}
\acro{BUT}{Brno University of Technology}
\acro{BDII}{BUT DIHARD II winning system}
\acro{BW}{Bandwidth}
\acro{BZ}{Back-to-Zero}
\acro{C4DM}{Centre for Digital Music}
\acro{C50}[$C_\textrm{50}$]{Clarity Index}
\acro{C-GFB}{Combination Gas-fired Boiler}
\acro{CART}{Classification and Regression Tree}
\acro{CASA}{Computational Auditory Scene Analysis}
\acro{CBR}{Constant Bit Rate}
\acro{CCC}{Cross-Correlation Coefficient}
\acro{CCCC}{DARPA CSR Corpus Coordinating Committee}
\acro{CCD}{Charge-Coupled Device}
\acro{CCI}{Call Clarity Index}
\acro{CCITT}{Consultative Committee for International Telephony and Telegraphy}
\acro{CCM}{Contralateral Competing Message}
\acro{CCR}{Comparison Category Rating}
\acro{CDB}{Constant Directivity Beamformer}
\acro{CDMA}{Code Division Multiple Access}
\acro{CELP}{Code-excited Linear Prediction}
\acro{CHIEF}{Combined Helmholtz Integral Equation Formulation}
\acro{CIT}{Constrained Initial Taps}
\acro{CL}{Clipping Level}
\acro{CLEAR}{Centre for Law Enforcement Audio Research}
\acro{CLID}{Cluster Identification Test}
\acro{CLT}{Central Limit Theorem}
\acro{CMA}{Constant Modulus Algorithm}
\acro{CMASI}{Coherence Modulated Acoustic Speckle Interferometry}
\acro{CMB}{Cosmic Microwave Background}
\acro{CMN}{cepstral mean normalisation}
\acro{CMVN}{cepstral mean and variance normalisation}
\acro{CNC}{Consonant-Nucleus-Consonant}
\acro{CNG}{Comfort Noise Generation}
\acro{CNN}{convolutional neural network}
\acro{CNNs}{convolutional neural networks}
\acro{CODEC}{Coder-Decoder}
\acro{COLA}{constant overlap-add}
\acro{CPE}{Customer Premises Equipment}
\acro{CPHD}{Cardinalized \acs{PHD}}
\acro{CPUs}{central processing units}
\acro{CRACD}{Codec-robust Automatic Clipping Detector}
\acro{CRC}{Cyclic Redundancy Check}
\acro{CS}{Channel Shortening}
\acro{CS2}[CS]{Compressive Sensing}
\acro{CSP}{Communications and Signal Processing}
\acro{CSR-WSJ}{Continuous Speech Recognition Wall Street Journal Phase 1\acroextra{ database}}
\acro{CST}{Connected Speech Test\acroextra{ speech corpus}}
\acro{CT}{Conversation Test}
\acro{CTS}{conversational telephone speech}
\acro{CTTN}{Comparative Tolerance to Noise}
\acro{CV}{Coefficient of Variation}
\acrodefplural{CV}{Coefficients of Variation}
\acro{CV2}[CV]{Constant Velocity}
\acro{CVC}{Consonant-Vowel-Consonant}
\acro{CW}{Continuous Wave}
\acro{CWM}{Centre-Weighted Median}
\acro{CWMY}{Centre-Weighted Myriad}
\acro{CWT}{continuous Wavelet transform}
\acro{DAC}{digital-to-analogue converter}
\acro{DAM}{Diagnostic Acceptability Measure}
\acro{DAQ}{Data Acquisition}
\acro{DARPA}{Defense Advanced Research Projects Agency\acroextra{ of the United States Dept. of Defense}}
\acro{DARPA-RMD}{\acs{DARPA} 1000-Word Resource Management Database\acroextra{ for Continuous Speech Recognition}}
\acro{DAW}{Digital Audio Workstation}
\acro{dB}{decibel}
\acro{dBFS}{\acs{dB} Full Scale}
\acro{DBN}{Deep Belief Network}
\acro{DBSTOI}{Deterministic \acs{BSTOI}}
\acro{DC}{Direct Current}
\acro{DCME}{Digital Circuit Multiplexing Equipment}
\acro{DCR}{Degradation Category Rating}
\acro{DCT}{discrete cosine transform}
\acro{2D-DCT}{two-dimensional discrete cosine transform}
\acro{ddCRP}{distance-dependent Chinese restaurant
process}
\acro{DDR}{Direct-to-diffuse ratio}
\acro{DDR3}{Double Data Rate Type Three}
\acro{DECT}{Digital European Cordless Telecommunication}
\acro{DeLILAH}{Detection of Clipping using Least Squares Residuals and Iterated Logarithm Amplitude Histogram}
\acro{DENBE}{\acs{DRR} Estimation using a Null-Steered Beamformer}
\acro{DER}{diarization error rate}
% \acro{DERe}[$DER_{\epsilon}$]{frame-based diarization error rate}
% \acro{DERt}[$DER_{\tau}$]{time-based diarization error rate}
\acro{DET}{Detection Error Trade-off}
\acro{Dev}{Development\acroextra{ dataset of the \acs{ACE} Challenge}}
\acro{DFT}{discrete Fourier transform}
\acro{DI}{Directivity Index}
\acro{DIHARD}{the DIHARD Speaker Diarization Challenge 2018 for hard speaker diarization problems}
\acro{DIHARD II}{the Second DIHARD Speaker Diarization Challenge 2019 for hard speaker diarization problems}
\acro{DIHARD III}{the Third DIHARD Speaker Diarization Challenge 2019 for hard speaker diarization problems}
\acro{DirectX}{\acroextra{A programming interface developed by Microsoft for handling tasks related to multimedia.}}
\acro{DIRHA}{Distant-speech Interaction for Robust Home Applications\acroextra{ multi-microphone multi-language acoustic speech corpus}}
\acro{DMA}{Differential Microphone Array}
\acro{DMT}{Discrete Multi-Tone}
\acro{DMV}{Dynamically Managed Voice\acroextra{ system}}
\acro{DNN}{deep neural network}
\acro{DNNs}{deep neural networks}
\acro{DNR}{Dynamic Noise Reduction}
\acro{DoA}{Direction-of-Arrival}
\acrodefplural{DoA}[DoAs]{Directions-of-Arrival}
\acro{DOA}{direction-of-arrival}
\acrodefplural{DOA}[DOAs]{Directions-of-Arrival}
\acro{DP}{Dynamic Programming}
\acro{DPCM}{Differential Pulse Code Modulation}
\acro{DPD}{Direct-Path Dominance}
\acro{DPD-MUSIC}{Direct-Path Dominance Multiple Signal Classification}
\acro{DR}{Douglas-Rachford}
\acro{DR2}{Dynamic Range}
\acro{DRM}{Diagnostic Rhyme Test}
\acro{DRR}{Direct-to-Reverberant Ratio}
\acro{DRNN}{Deep Recurrent Neural Net}
\acro{DRT}{Diagnostic Rhyme Test}
\acro{DSB}{Delay-and-Sum Beamformer}
\acro{DSOBM}{Deterministic \acs{SOBM}}
\acro{DSP}{digital signal processing}
\acro{DSPS}{Double Sides Periodic Substitution}
\acro{DSR}{Distributed Speech Recognition}
\acro{DSWS}{Double Sides Waveform Substitution}
\acro{DTFT}{discrete time Fourier transform}
\acro{DTMF}{Dual Tone Multi-Frequency}
\acro{DTX}{Discontinued Transmission}
\acro{DWT}{discrete wavelet transform}
\acrodefplural{DWT}[DWTs]{discrete wavelet transforms}
\acro{e2e}{end-to-end}
\acro{EARS}{Embodied Audition for RobotS}
\acro{EBF}{Eigen-beamformer}
\acro{EBU}{European Broadcasting Union}
\acro{EC}{Echo Canceller}
\acro{EDC}{Energy Decay Curve}
\acro{EDR}{Energy Decay Relief}
\acro{EDF}{Energy Decay Function}
\acro{EEG}{Electroencephalography}
\acro{EER}{Equal Error Rate}
\acro{EFICA}{Efficient Fast Independent Component Analysis}
\acro{EIR}{Equalized Impulse Response}
\acro{EKF}{Extended Kalman Filter}
\acro{EKF-SLAM}[EKF-SLAM]{\acs{EKF} \acs{SLAM}}
\acro{EKF-SLAM2}[EKF-SLAM]{Extended Kalman Filter \acs{SLAM}}
\acro{EL}{Echo Loss}
\acro{ELBO}{evidence lower bound}
\acro{ELF}{Extremely Low Frequency}
\acro{ELRA}{European Languages Research Association}
\acro{EM}{Estimation-Maximization\acroextra{. An iterative technique to solve certain optimization problems.}}
\acro{EMD}{empirical mode decomposition}
\acro{EEMD}{ensemble empirical mode decomposition}
\acro{EMIB}{Eigenmike Microphone Interface Box}
\acro{ENF}{Electrical Network Frequency}
\acro{ENV}{temporal envelope}
\acro{EPSRC}{Engineering and Physical Sciences Research Council}
\acro{EQ}{Equalisation}
\acro{ERB}{Equivalent Rectangular Bandwidth}
\acro{ERP}{Ear Reference Point (cf. ITU-T Rec. P.64 1999)}
\acro{ESA}{Early Stage Assessment}
\acro{ESM}{Equivalent Source Method}
\acro{ESPRIT}{Estimation of Signal Parameters via Rotational Invariance Techniques}
\acro{ETAN}{Equivalent Tolerance to Additional Noise} 
\acro{ETSI}{European Telecommunications Standards Institute}
\acro{EURASIP}{European Association for Signal Processing}
\acro{EUSIPCO}{European Signal Processing Conference}
\acro{Eval}{Evaluation\acroextra{ dataset of the \acs{ACE} Challenge}}
\acro{F0}{the fundamental frequency}
\acro{F1}{F1 Score}
\acro{FAR}{False Alarm Rate}
\acro{FastSLAM}[FastSLAM]{FActored Solution To Simultaneous Localization and Mapping}
\acro{FastSLAM2}[FastSLAM]{FActored Solution To \acs{SLAM}} 
\acro{FAU}{Friedrich-Alexander-Universit{\"a}t}
\acro{FB}{Forward-Backward}
\acro{FB2}[FB]{Fullband}
\acro{FBANKs}{mel filter bank cepstral coefficients}
\acro{FBCCs}{mel filter bank cepstral coefficients}
\acro{FBF}{Fixed Beamformer}
\acro{FBSS}{Forward-Backward Spatial Smoothing}
\acro{FCC}{Federal Communications Commission}
\acro{FDLP}{frequency domain linear prediction}
\acro{FDM}{Frequency Division Multiplexing}
\acro{FDR}{False Discovery Rate}
\acro{FDR2}[FDR]{Free-Decay Region}
\acro{FEC}{Forward Error Correction}
\acro{FEM}{Finite Element Method}
\acro{FFI}{Norwegian Defence Research Establishment}
% \acro{FFT}{Fast Fourier Transform}\acroindefinite{FFT}{an}{a} % will break old versions of Lyx preamble
\acro{FFT}{fast Fourier transform}
\acroindefinite{FFT}{an}{a}
\acro{FIFO}{first-in first-out}
\acro{FIR}{finite impulse response\acroextra{. A filter whose output is a weighted sum of past input values and whose system function contains only zeros and no poles.}}
\acro{FISM}{Fast Image Source Method}
\acro{FISST}{Finite Set STatistics}
\acro{FLOM}{Fractional Lower-Order Moments}
\acro{FLOS}{Fractional Lower-Order Statistics}
\acro{FM}{frequency modulation}
\acro{FMM}{Fast Multipole Method}
\acro{FN}{False Negative}
\acro{FN2}{Nth Formant}
\acro{FNR}{False Negative Rate}
\acro{FORTRAN}{The IBM Mathematical Formula Translating System}
\acro{FoV}{Field of View}
\acro{FP}{False Positive}
\acro{FPR}{False Positive Rate}
\acro{FPS}{Frames Per Second}
\acro{FRI}{Finite Rate of Innovation}
\acro{FSB}{Filter-and-Sum Beamformer}
\acro{FSK}{Frequency Shift Keying}
\acro{FT}{Flat-Top}
\acro{FWER}{Familywise Error Rate}
\acro{FWSSNR}{Frequency-Weighted Segmental \acs{SNR}}
\acro{FWSSRR}{Frequency-Weighted \acs{SSRR}}
\acro{G.711}{\acs{PCM} of Voice Frequencies}
\acro{GARCH}{Generalized Auto-regressive Conditional Heteroscedasticity}
\acro{GBW}{Gain Bandwidth Product}
\acro{GCC}{generalized cross-correlation}
\acro{GCC-PHAT}{generalized cross-correlation with phase transform\acroextra{ method of estimating \acs{TDOA}}}
\acro{GCF}{Global Coherence Field}
\acro{GGD}{Generalized Gaussian Distribution}
\acro{GGD2}[G$\Gamma$D]{Generalised Gamma Distribution}
\acro{GLR}{generalised likelihood ratio}
\acro{GM}{Gaussian mixture}
\acro{GM-PHD}{Gaussian Mixture \acs{PHD}}
\acro{GMCA}{Generalized Morphological Component Analysis}
%\acro{GMM}{Gaussian mixture model\acroextra{. An approximation to an arbitrary probability density function that consists of a weighted sum of Gaussian distributions}}
\acro{GMM}{Gaussian mixture model}
\acro{GNSS}{Global Navigation Satellite System}
\acro{GPRS}{General Packet Radio Services}
\acro{GPS}{Global Positioning System}
\acro{GPUs}{graphics processing units}
\acro{GSC}{Generalized Sidelobe Canceller}
\acro{GSM}{Global System for Mobile Communications}
\acro{GSM-EFR}{\acs{GSM} Enhanced Full Rate Codec}
\acro{GSM-FR}{\acs{GSM} Full Rate Codec}
\acro{GSM-HR}{\acs{GSM} Half Rate Codec}
\acro{GT}{ground truth}
\acro{GT1}{ground truth labels based on ...}
\acro{GT2}{ground truth labels based on ...}
\acro{GT3}{ground truth labels based on ...}
\acro{GT4}{ground truth labels based on ...}
\acro{GT-labels}{ground truth labels}
\acro{GT-SAD}{ground truth speech activity detection}
\acro{GUI}{Graphical User Interface}
\acro{HAAC}{High Amplitude Audio Capture}
\acro{HATS}{Head and Torso Simulator}
\acro{HERB}{Harmonicity-based dEReverBeration}
\acro{HFT}{Hands-Free Terminal}
\acro{HHT}{Hilbert-Huang transform}
\acro{HI}{Hearing-Impaired}
\acro{HIE}{Helmholtz Integral Equation}
\acro{HISAS}{High resolution Interferometric \ac{SAS}}
\acro{HINT}{Hearing-in-Noise Test}
\acro{HLT}{Human Language Technology}
\acro{HMM}{hidden Markov model}
\acro{HOS}{Higher-Order Statistics}
\acro{HPF}{High-Pass Filter}
\acro{HR}{Half Rate (\acs{GSM} Codec)}
\acro{HRI}{Human-Robot Interaction}
\acro{HRTF}{Head-related Transfer Function}
\acro{HSA}{Hearing, Speech, Audio\acroextra{ technology group at Fraunhofer IDMT}}
\acro{HSD}{Hybrid Steepest Descent}
\acro{HT}{Hannan-Thomson}
\acro{HTK}{Hidden Markov Model Tool Kit}
\acro{IBM}{Ideal Binary Mask}
\acro{IC}{Interference Canceller}
\acro{ICA}{Independent Component Analysis}
\acro{ICASSP}{Intl. Conf. on Acoustics, Speech and Signal Processing}
\acro{ICSI}{International Computer Science Institute}
\acro{ID}{Identifier}
\acro{Idiap}{Idiap Research Institute}
\acro{IDMT}{Institute for Digital Media Technology}
\acro{iDEN}{Integrated Digital Enhanced Network}
\acro{IDTFT}{inverse discrete time Fourier transform}
\acro{IEC}{International Electrotechnical Commission}
\acro{IEEE}{Institute of Electrical and Electronics Engineers}
\acro{IET}{Institute of Engineering and Technology}
\acro{IETF}{Internet Engineering Task Force}
\acro{IFCs}{instantaneous frequency coefficients}
\acro{IFCCs}{instantaneous frequency cosine coefficients}
\acro{IFFT}{inverse fast Fourier transform}
\acro{IHC}{Inner Hair Cell}
\acro{IHT}{Iterative Hard Thresholding}
\acro{IID}{Independent and Identically Distributed}
\acro{IIR}{Infinite Impulse Response\acroextra{. A filter whose output is a weighted sum of both past input and past output values and whose system function contains both poles and zeros.}}
\acro{IL}{Iterated Logarithm\acroextra{, the logarithm of the logarithm}}
\acro{ILAH}{Iterated Logarithm Amplitude Histogram\acroextra{ clipping detection method}}
\acro{ILD}{Interaural Level Difference}
\acro{IMCRA}{Improved Minima Controlled Recursive Averaging\acroextra{. A technique for blindly estimating the spectrum of additive noise in a signal.}}
\acro{IMD}{Inter-Modulation Distortion}
\acro{IMF}{intrinsic mode function}
\acro{IMSI}{International Mobile Subscriber Identity}
\acro{IMU}{Inertial Measurement Unit}
\acro{INMD}{In-service Non-intrusive Measurement Device}
\acro{INTERSPEECH}{Annual Conference of the \acs{ISCA}}
\acro{IO}{Infinitely Often}
\acro{IP}{Internet Protocol}
\acro{IPA}{International Phonetic Association}
\acro{IPD}{Interaural Phase Difference}
\acro{IRM}{Ideal Ratio Mask}
\acro{IRS}{Inverse repeated Sequence\acroextra{. A pseudo random sequence used for impulse response measurement.}}
\acro{IRS2}[IRS]{Intermediate Reference System}
\acro{IS}{Importance Sampling}
\acro{ISCA}{International Speech Communication Association}
\acro{ISDN}{Integrated Services Digital Network}
\acro{ISFT}{Inverse \acl{SFT}}
\acro{ISO}{Intl. Organization for Standardization}
\acro{ISP}{Intensity Spectral Profile}
\acro{IST}{Iterative Soft Thresholding}
\acro{ISTFT}{inverse short-time discrete Fourier transform}
\acro{ISVR}{Institute of Sound and Vibration Research\acroextra{, Southampton University, UK}}
\acro{ITD}{Interaural Time Difference}
\acro{ITF}{Interaural Transfer Function}
\acro{ITU}{International Telecommunication Union}
\acro{ITUR}[ITU-R]{International Telecommunication Union\acroextra{ Radiocommunication Sector}}
\acro{ITUT}[ITU-T]{International Telecommunication Union\acroextra{ Telecommunication Standardisation Sector}}
\acro{IUWT}{Isotropic Undecimated Wavelet (Starlet) Transform}
\acro{IWAENC}{Intl. Workshop on Acoustic Signal Enhancement}
\acro{IWAENC_PRE_2012}[IWAENC]{Intl. Workshop Acoustic Echo and Noise Control}
\acro{IWASE}{Intl. Workshop on Acoustic Signal Enhancement}
\acro{JADE}{Joint Approximate Diagonalization of Eigen-Matrices}
\acro{JER}{Jaccard Error Rate}
\acro{JFA}{joint factor analysis}
\acro{JHU}{Johns Hopkins University}
\acro{JPDA}{Joint Probabilistic Data Association}
\acro{JPEG}{Joint Photographic Experts Group}
\acro{JSD}{Jensen-Shannon divergence}
\acro{JSID}{joint speaker diarization and identification}
\acro{KDE}{Kernel Density Estimate}
\acro{KF}{Kalman Filter}
\acro{KFD}{Kernel Fisher Discriminant\acroextra{ Analysis}}
%\acro{KL}{Karhunen-Lo{\'{e}}ve}
\acro{KLD}{Kullback-Leibler divergence}
\acro{KLT}{Karhunen-Lo{\'{e}}ve transform}
\acro{KST}{Kolmogorov-Smirnov Test}
\acro{LAD}{Least Absolute Deviation}
\acro{LAN}{Local Area Network}
\acro{LARS}{Least Angle Regression}
\acro{LAT}[L$_{\textrm{AT}}$]{Equivalent Continuous Sound Level\acroextra{. Also called Leq}}
\acro{LBR}{Low Bitrate Redundancy}
\acro{LC}{Local Criterion}
\acro{LCMP}{Linearly Constrained Minimum Power}
\acro{LCMV}{Linearly Constrained Minimum Variance}
\acro{LCWM}{Linear Combination of Weighted Medians}
\acro{LDA}{linear discriminant analysis}
\acro{LDC}{Linguistic Data Consortium}
\acro{LEM}{Loudspeaker-Enclosure-Microphone System}
\acro{Leq}[L$_{\textrm{eq}}$]{Equivalent Continuous Sound Level\acroextra{. Also called LAT}}
\acro{LF}{Liljencrants-Fant\acroextra{. The developers of a glottal waveform model}}
\acro{LHS}{Left-Hand Side}
\acro{LID}{Language Identification}
\acro{LILAH}{\acs{LSR2}-\acs{ILAH}\acroextra{ clipping detection method}}
\acro{LIME}{LInear Predictive Multi-input Equalization\acroextra{ algorithm}}
\acro{LiNoPS}{Lightweight Noise Protection System}
\acro{LLN}{Law of Large Numbers}
\acro{LLR}{log-likelihood ratio}
\acro{LLS}{Logarithmic Least Squares}
\acro{LMA}{Least Mean Absolute}
\acro{LMS}{least mean squares\acroextra{ adaptive filter}}
\acro{LNA}{Low Noise Amplifier}
\acro{LoS}{Line-of-Sight}
\acro{LOT}{Listening-Only Test}
\acro{LP}{Linear Parameter}
\acro{LP2}[LP]{Linear Predictive}
\acro{LP3}[LP]{Linear Prediction}
%\acro{LPC}{linear predictive coding\acroextra{. An autoregressive model of speech production.}}
\acro{LPC}{linear predictive coding}
\acro{LPCCs}{linear predictive cepstral coefficients}
\acro{LQO}{Listening Quality Objective}
\acro{LR}{likelihood ratio}
\acro{LREC}{Conf. on Language Resources and Evaluation}
\acro{LS}{Least Squares}
\acro{LSA}{Log Spectral Amplitude}
\acro{LSB}{Lower Side-Band}
\acro{LSB2}[LSB]{Least Significant Bit}
\acro{LSD}{Log Spectral Distortion}
\acro{LSFs}{line spectral frequencies}
%\acro{LSP}{Line Spectrum Pairs}
\acro{LSPs}{line spectral pairs}
\acro{LSR}{Late Stage Review}
\acro{LSR2}[LSR]{Least Squares Residuals\acroextra{ clipping detection method}}
\acro{LSRT}{Least Squares Residuals with Thresholding\acroextra{ clipping detection method}}
\acro{LSTM}{recurrent neural network - long short-term memory}
\acro{LTASS}{Long Term Average Speech Spectrum}
\acro{LTI}{Linear Time Invariant}
\acro{LTP}{Long Term Prediction}
\acro{LU}{Loudness Unit}
\acro{LUFS}{Loudness Units Full-Scale}
\acro{MA}{Moving Average}
\acro{MAC}{Multiply Accumulate Operation}
\acro{MAD}{Median Absolute Deviation}
\acro{MAE}{Mean Absolute Error}
\acro{MAP}{maximum \emph{a posteriori}}
\acro{MARDY}{Multichannel Acoustic Reverberation Database at York}
\acro{MARS}{Multivariate Adaptive Regression Splines}
\acro{MBF}{Matched Filter Beamformer}
\acro{MC}{Monte Carlo}
\acro{MCA}{Morphological Component Analysis}
\acro{MCC}{Matthew's Correlation Coefficient}
\acro{MC Dropout}{Monte Carlo dropout}
\acro{MCEQ}{MultiChannel EQualisation}
\acro{MCS}{Multidimensional Colouration Space}
\acro{MDCCs}{modulation domain cepstral coefficients}
\acro{MDCT}{Modified Discrete Cosine Transform}
\acro{MDL}{Minimum Description Length}
\acro{MDM}{multiple distant microphones}
\acro{MCMC}{Markov chain Monte Carlo}
\acro{MDS}{Multidimensional Scaling}
\acro{Mel}{\acroextra{A non-uniform frequency scale corresponding to perceived frequency. It is approximately linear at low frequencies and logarithmic at high frequencies.}}
\acro{MELP}{Mixed Excitation Linear Prediction}
\acro{MEMS}{micro electro-mechanical systems}
\acro{MFCCs}{mel-frequency cepstral coefficients}
\acro{MFS}{Method of Fundamental Solutions}
\acro{MFSK}{Multi-Frequency Shift Keying}
\acro{MHT}{Multi-Hypotheses Tracking}
\acro{MI}{Mutual Information}
\acro{MIMO}{Multiple-Input-Multiple-Output}
\acro{MINT}{Multiple-input/output INverse Theorem}
\acro{MIRS}{Motorola Integrated Radio System}
\acro{MIT}{Massachusetts Institute of Technology}
\acro{MIT-LCS}{Massacchusetts Institute of Technology Laboratory for Computer Science}
\acro{ML}{maximum likelihood}
\acro{MLE}{maximum likelihood estimation}
\acro{ML-TDoA}{Maximum Likelihood Time Difference of Arrival}
\acro{MLD}{Masking Level Difference}
\acro{MLLR}{maximum likelihood linear regression}
\acro{MLMF}{Machine Learning with Multiple Features}
\acro{MLS}{Maximum Length Sequence\acroextra{ of pseudo random bits.}}
\acro{MMSE}{Minimum Mean Squared Error}
\acro{MMT}{Multiscale Median Transform}
\acro{MNRU}{Modulated Noise Reference Unit}
\acro{MOM}{Mean of Maximum}
\acro{MOS}{Mean Opinion Score}
\acro{MOS-LQO}{Mean Opinion Score - Listening Quality Objective}
\acro{MP}{Matching Pursuit}
\acro{MP3}{\acs{MPEG}-2 Audio Layer III}
\acro{MPEG}{Moving Picture Experts Group}
\acro{MRF}{Markov Random Field}
\acro{MRP}{Mouth Reference Point (cf. ITU-T Rec. P.64 1999)}
\acro{MRT}{Modified Rhyme Test}
\acro{MS}{Minimum Statistics}
\acro{MSB}{Most Significant Bit}
\acro{MSC}{Mean Square Coherence}
\acro{MSE}{Mean Square Error}
\acro{MSG}{modulation spectrogram}
\acro{MSN}{Multiple Subscriber Number}
\acro{MSNR}{Maximum \acs{SNR}}
\acro{MTF}{Modulation Transfer Function}
\acro{MTM}{Modified Trimmed Mean}
\acro{MUSCLE}{MeasUred Single-CLustEr}
\acro{MUSHRA}{Multi-stimuli Test with Hidden Reference and Anchor}
\acro{MUSHRAR}{Multi-stimuli Test with Hidden Reference and Anchor for Reverberation}
\acro{MUSIC}{Multiple Signal Classification}
\acro{MVDR}{Minimum Variance Distortionless Response\acroextra{ beamformer}}
\acro{MWF}{Multi-channel Wiener Filter}
\acro{NAH}{Nearfield Acoustic Holography}
\acro{NB}{Narrowband}
\acro{NCM}{Normalized Coherence Metric}
\acro{NH}{Normal-Hearing}
\acro{NIRA}{Non-Intrusive Room Acoustics}
\acro{NISE}{Non-Intrusive \acs{SNR} estimation}
\acro{NISI}{Non-Intrusive Speech Intelligibility Estimation}
\acro{NISQ}{Non-Intrusive Speech Quality Estimation}
\acro{NIST}{National Institute of Standards and Technology}
\acro{NL}{Noise Level}
\acro{NLA}{Non-Linear Approximation}
\acro{NLL}{negative log-likelihood}
\acro{NLP}{natural language processing}
\acro{NLMS}{Normalized Least Mean Squares\acroextra{adaptive filter}}
\acro{NMCFLMS}{Normalized Multichannel Frequency Domain Least Mean Square}
\acro{NMF}{Non-negative Matrix Factorization}
\acro{NMI}{normalized mutual information}
\acro{NOISEX-92}{Database to Study the Effect of Additive Noise on Speech Recognition Systems}
\acro{NOIZEUS}{Noisy Speech Corpus for Evaluation of Speech Enhancement Algorithms}
\acro{NOSRMR}{Normalized Overall \acs{SRMR}}
\acro{NOS}{Number of Sources}
\acro{NOSRMR}{Normalized Overall \acs{SRMR}}
\acro{NPM}{Normalized Projection Misalignment}
\acro{NPV}{Negative Predictive Value}
\acro{NR}{Noise Reduction}
\acro{NS}{Noise Suppression}
\acro{NSRMR}{Normalised \acs{SRMR}}
\acro{NSRR}{Normalized Signal-to-Reverberation Ratio}
\acro{NSRMR}{Normalised \acs{SRMR}}
\acro{NSV}{Negative-Side Variance}
\acro{NTP}{Network Time Protocol}
\acro{NV}{Noise-Vocoding}
\acro{OBL}{Octave Band Level}
\acro{ODF}{Overdrive Factor}
\acro{Ofcom}{Office of Communications\acroextra{, the independent regulator and competition authority for the UK communications industries}}
\acro{OFDM}{Orthogonal Frequency Division Multiplexing}
\acro{OHC}{Outer Hair Cell}
\acro{OIM}{Objective Intelligibility Measure}
\acro{OLA}{Overlap-add}
\acro{OM-LSA}{Optimally Modified Log-Spectral Amplitude{ Estimator}}
\acro{OMP}{Orthogonal Matching Pursuit}
\acro{OSI}{Open Systems Interconnection}
\acro{OSPA}{Optimal Subpattern Assignment}
\acro{OSRMR}{Overall \acs{SRMR}}
\acro{OSTI-SI}{open-set text-independent speaker identification}
\acro{PAB-SRMR}{Per acoustic band \acs{SRMR}}
\acro{PALM}{Passive Acoustic Localization and Mapping}
\acro{PAMS}{Perceptual Analysis Measurement System}
\acro{PARCOR}{Partial Correlation Coefficients}
\acro{PB}{Phonetically Balanced}
\acro{PBF}{Positive Boolean Function}
\acro{PCA}{principal component analysis}
\acro{PCM}{pulse code modulation}
\acro{PDA}{Personal Digital Assistant}
\acro{PDA2}[PDA]{Probabilistic Data Association}
\acro{PDE}{partial differential equation}
\acro{pdf}{probability density function}
\acro{PDF}{probability density function}
\acro{PE}{Parameter Estimation}
\acro{PEASS}{Perceptual Evaluation for Audio Source Separation}
\acro{PEFAC}{Pitch Estimation Filter with Amplitude Compression}
\acro{PESQ}{Perceptual Evaluation of Speech Quality}
\acro{PF}{Psychometric Function}
\acro{pgfl}[p.g.fl.]{Probability Generating Functional}
\acro{PLDA}{probabilistic linear discriminant analysis}
\acro{PHAT}{Phase Transform}
\acro{PHD}{Probability Hypothesis Density}
\acro{PIP}{Peak-Image Pairing}
\acro{PIV}{Pseudo-Intensity Vector}
\acro{PL}{Pseudo-likelihood}
\acro{PLC}{Packet Loss Concealment}
\acro{PLL}{Phase Locked Loop}
\acro{PLP}{perceptual linear prediction}
\acro{PLPCCs}{perceptual linear prediction cepstral coefficients}
\acro{PLR}{Perceived Level of Reverberation}
\acro{PM}{phase modulation}
\acro{PMF}{Probability Mass Function}
\acro{PMOS}{Predicted Mean Opinion Score}
\acro{POLQA}{Perceptual Objective Listening Quality Analysis}
\acro{POTS}{Plain Old Telephone Service}
\acro{PPCA}{probabilistic principal component analysis}
\acro{PPP}{Poisson Point Process}
\acro{PPS}{Pulse-Per-Second}
\acro{PPV}{Positive Predictive Value}
\acro{PRLM}{Phoneme Recognition and Language Modelling}
\acro{PRP}{Pair-wise Relative Phase-ratio}
\acro{PSD}{Power Spectral Density}
\acro{PSK}{Phase Shift Keying}
\acro{PSNR}{Peak Signal-to-Noise Ratio}
\acro{PSOLA}{Pitch Synchronous Overlap Add\acroextra{. A method of scaling a signal in time and pitch independently.}}
\acro{PSQM}{Perceptual Speech Quality Measurement}
\acro{PSTN}{Public Switched Telephone Network}
\acro{PWD}{Plane-Wave Decomposition}
\acro{PTA}{Pure-Tone Audiology}
\acro{QAM}{Quadrature Amplitude Modulation}
\acro{QILAH}{Quadrisected Iterated Logarithm Amplitude Histogram\acroextra{ clipping detection method}}
\acro{QMF}{Quadrature Mirror Filter}
\acro{QoE}{Quality-of-Experience}
\acro{QoS}{Quality-of-Service}
\acro{QPSK}{Quadrature Phase Shift Keying}
\acro{RADAR}{RAdio Detection And Ranging}
\acro{RASTA}{relative spectral}
\acro{RASTA-PLP}{relative spectral perceptual linear prediction}
\acro{RASTA-PLPCCs}{relative spectral perceptual linear prediction cepstral coefficients}
\acro{RASTI}{Room Acoustics Speech Transmission Index\acroextra{ (superseded by STIPA)}}
\acro{RBM}{Restricted Boltzmann Machine}
\acro{RB-PHD}{Rao-Blackwellised \acs{PHD}}
\acro{RBPF}{Rao-Blackwellised Particle Filter}
\acro{RC}{Relative Criterion}
\acro{RDT}[$R_\textrm{DT}$]{Reverberation Decay Tail}
\acro{RDTF}{Relative Direct Transfer Function}
\acro{ReLU}{rectified linear unit}
\acro{RF}{Radio Frequency}
\acro{RFI}{Radio Frequency Interference}
\acro{RFS}{Random Finite Set}
\acro{RHS}{Right-Hand Side}
\acro{RIP}{Restricted Isometry Property}
\acro{RIR}{Room Impulse Response}
\acro{RLS}{Recursive Least Squares\acroextra{ adaptive filter}}
\acro{RLSD}{Relative Log Spectral Distortion}
\acro{RMCLS}{Relaxed Multichannel Least Squares}
\acro{RMCLS-CIT}{Relaxed MultiChannel Least-Squares with Constrained Initial Taps}
\acro{RMS}{Root Mean Square}
\acro{RMSE}{Root Mean Square Error}
\acro{RNID}{Royal National Institute for Deaf People}
\acro{RNN}{recurrent neural network}
\acro{RNNs}{recurrent neural networks}
\acro{RNN-LSTM}{recurrent neural network - long short-term memory}
\acro{RNN-LSTMs}{recurrent neural networks - long short-term memory}
\acro{ROC}{Receiver Operating Characteristic}
\acro{ROHC}{Robust Header Compression}
\acro{RP}{Received Pronunciation}
\acro{RPE}{Regular Pulse Excitation}
\acro{RRTF}{Relative Real-Time Factor}
\acro{RS}{Reverberation Suppression}
\acro{RSM}{Reflector Source Method}
\acro{RSV}{Room Spectral Variance}
\acro{RT}{Reverberation Time}
\acro{RT-05S}{Spring 2005 (RT-05S) NIST Rich Transcription Meeting Evaluation}
\acro{RTAN}{Robustness to Additional Noise}
\acro{RTF}{Real-Time Factor}
\acro{RTF2}[RTF]{Room Transfer Function}
\acro{RTF3}[RTF]{Relative Transfer Function}
\acro{RTS}{Rauch-Tung-Striebel}
\acro{RTTM}{rich transcription time marked}
\acro{RV}{Random Variable}
\acro{RVP}{Recursive Vector Projection}
\acro{RWTH}{Rheinisch-Westf\"{a}lische Technische Hochschule}
\acro{S50}[$S_{50}$]{Intelligibility Function Gradient at the \ac{SRT}}
\acro{SA}{Spectral Amplitude}
\acro{SAA}{Synthetic Aperture Audio}
\acro{SAD}{speech activity detection}
\acro{SAP}{Speech And Audio Processing}
\acro{SAR}{Speech-to-Artifact Ratio}
\acro{SAR2}[SAR]{Speaker Alternation Rate}
\acro{SAR3}[SAR]{Synthetic Aperture \ac{RADAR}}
\acro{SAS}{Synthetic Aperture \ac{SONAR}}
\acro{SB}{Subband}
\acro{SBDER}{segment boundary diarization error rate}
\acro{SC-PHD}{Single Cluster \acs{PHD}}
\acro{SCAF}{Single Channel Adaptive Filter}
\acro{SCB}{Stochastic Codebook}
\acro{SCD}{speaker change detection}
\acro{SCNR}{Single-Channel Noise Reduction}
\acro{SCOT}{Smoothed Coherence Transform}
\acro{SCNR}{Single-Channel Noise Reduction}
\acro{SC-PHD}{Single Cluster \acs{PHD}}
\acro{SCR}{Signal-to-Competition Ratio}
\acro{SCRIBE}{Spoken Corpus of British English}
\acro{SCT}{Speech Corruption Toolkit}
\acro{SCT2}{Short Conversation Test}
\acro{SD}{Semantic Differential}
\acro{SDB}{Superdirective Beamformer}
\acro{SDD}{Spectral Decay Distributions}
\acro{SDDMSB}{\acs{SDD} with Mel-spaced frequency bands}
\acro{SDDSA}{\acs{SDD} with Mel-spaced frequency bands and selective averaging}
\acro{SDDSA-G}{\acs{SDDSA} with Gerkmann noise estimator}
\acro{SDDSA-H}{\acs{SDDSA} with Hendriks noise estimator}
\acro{SDM}{single distant microphone}
\acro{SDR}{Software Defined Radio}
\acro{SDR2}{Speech}
\acro{SDRAM}{Synchronous Dynamic Random Access Memory}
\acro{SDT}{Speech Description Taxonomy}
\acro{SEDF}{Subband \ac{EDF}}
\acro{SEMG}{Surface Electromyography}
\acro{SFDR}{Spurious Free Dynamic Range}
\acro{SFM}{Single Feature with Mapping}
\acro{SFT}{Spherical Fourier Transform}
\acro{SH}{Spherical Harmonic}
\acro{SHD}{Spectral Harmonic Decomposition}
\acro{SHD2}[SHD]{Spherical Harmonic Domain}
\acro{sIB}{sequential information bottleneck}
\acro{SIE}{System Identification Error}
\acro{SII}{Speech Intelligibility Index}
\acro{SIImod}{Speech Intelligibility Index in the modulation domain}
\acro{SIM}{Subscriber Identity Module}
\acro{SIMO}{Single-Input-Multiple-Output}
\acro{SINAD}{Signal-to-Noise and Distortion Ratio}
\acro{SIP}{Session Initiation Protocol}
\acro{SIR}{Signal-to-Interference Ratio}
\acro{SIR2}[SIR]{Sequential Importance Resampling}
\acro{SIREAC}{Simulation of REal Acoustics\acroextra{ Software Tool}}
\acro{SIS}{Sequential Importance Sampling}
\acro{SL}{Speech Level}
\acro{SLAM}{Simultaneous Localization and Mapping}
\acro{SLLN}{Strong Law of Large Numbers}
\acro{SLM}{Sound Level Meter}
\acro{SMA}{Spherical Microphone Array}
\acro{SMARD}{Single- and Multichannel Audio Recordings Database}
\acro{SMERSH}{Spatiotemporal Averaging Method for Enhancement of Reverberant Speech}
\acro{SMIR}{Spherical Microphone array Impulse Response}
\acro{SMPTE}{Society of Motion Picture and Television Engineers}
\acro{SMS}{Short Message Service}
\acro{SNR}{signal-to-noise ratio}
\acro{SNR2}[SNR]{Speech-to-Noise Ratio}
\acro{SNT}{Subspace Noise Tracking\acroextra{ algorithm}}
\acro{SOBM}{STOI-optimal Binary Mask}
\acro{SONAR}{SOund Navigation And Ranging}
\acro{SOLA}{Synchronous Overlap Add\acroextra{. A method of scaling a signal in time and pitch independently.}}
\acro{SPC}{Specificity}
\acro{SPEECON}{Speech Databases for Consumer Devices}
\acro{SPHERE}{NIST SPeech Header REsources\acroextra{ software with embedded Shorten Compression}}
\acro{SPIN}{Speech Perception In Noise}
\acro{SPL}{Sound Pressure Level}
\acro{SPP}{Speech Presence Probability}
\acro{SPQA}{Speech Quality Assurance Package}
\acro{SQNR}{Signal-to-Quantization Noise Ratio}
\acro{SR}{Sparse Representation}
\acro{SR2}[SR]{Spectral Rotation}
\acro{SRA}{Statistical Room Acoustics}
\acro{SRI}{SRI International\acroextra{. Formerly Standford Research Institute}}
\acro{SRMR}{Speech-to-Reverberation Modulation Energy Ratio}
\acro{SRP}{steered response power}
\acro{SRP-PHAT}{steered response power with phase transform}
\acro{SRP-TDE}{Steered Response Power with Time Delay Estimation}
\acro{SRR}{Signal-to-Reverberation Ratio}
\acro{SRT}{Speech Reception Threshold\acroextra{ (also known as Speech Recognition Threshold)}}
\acro{SS}{Spectral Subtraction}
\acro{SSB}{Single Side-Band}
\acro{SSI}{Synthetic Sentence Identification}
\acro{SSL}{Sound Source Localization}
\acro{SSN2}[SSN]{Simultaneous Switching Noise}
\acro{SSN}{Speech-Shaped Noise}
\acro{SSNR}{Segmental \acs{SNR}}
\acro{SSOBM}{Stochastic \acs{SOBM}}
\acro{SSRR}{Segmental Signal-to-Reverberation Ratio}
\acro{SSW}{Staggered Spondaic Word}
\acro{STFT}{short-time discrete Fourier transform}
\acro{STI}{Speech Transmission Index}
\acro{STIPA}{Speech Transmission Index for Public Address Systems}
\acro{STITEL}{Speech Transmission Index for Telecommunication Systems}
\acro{STMI}{Spectro-Temporal Modulation Index}
\acro{STNR}{\acs{NIST}'s Speech-to-Noise Ratio\acroextra{ Estimation Algorithm}}
\acro{STOI}{short-time objective intelligibility\acroextra{ measure}}
\acro{STQ}{Speech Processing, Transmission and Quality Aspects}
\acro{STSA}{Short Time Spectral Analysis}
\acro{STSA1}[STSA]{Short Time Spectral Amplitude}
\acro{SUS}{Semantically Unpredictable Sentences}
\acro{SVD}{singular value decomposition}
\acro{SVM}{Support Vector Machine}
\acro{SWSOBM}{Stochastic \acs{WSOBM}}
\acro{T20}[$T_\textrm{20}$]{Reverberation Time\acroextra{ to decay by $20$ dB}}
\acro{T30}[$T_\textrm{30}$]{Reverberation Time\acroextra{ to decay by $30$ dB}}
\acro{T60}[$T_\textrm{60}$]{Reverberation Time\acroextra{ to decay by $60$ dB}}
\acro{TBM}{Target Binary Mask}
\acro{TDE}{Time Delay Estimation}
\acro{TDHS}{Time Domain Harmonic Scaling\acroextra{. A method of scaling a signal in time and pitch independently.}}
\acrodefplural{TDOA}[TDOAs]{time differences of arrival}
\acro{TDNN}{time delay neural network}
\acro{TDOA}{time difference of arrival}
\acrodefplural{TDOA}[TDOAs]{time differences of arrival}
\acro{TDT}{Tone Decay Test}
\acro{TF}{time-frequency}
\acro{TFS}{temporal fine structure}
\acro{TFGM}{Time-Frequency Gain Modification\acroextra{. An approach to signal enhancement in which a signal is multiplied by a gain function in the time-frequency domain.}}
\acro{THD}{Total Harmonic Distortion}
\acro{TI}{Texas Instruments, Inc.}
\acro{TIMIT}{\acs{TI}-\acs{MIT} speech corpus}
\acro{TIPHON}{Telecommunication and Internet Protocol Harmonization Over Networks}
\acro{TLS}{Total Least-Squares}
\acro{TN}{True Negative}
\acro{TNR}{True Negative Rate}
\acro{TOA}{time of arrival}
\acrodefplural{TOA}[TOAs]{times of arrival}
\acro{TOF}{Time-of-Flight}
\acro{TOSQA}{Telekom Objective Speech Quality Assessmentt}
\acro{TP}{True Positive}
\acro{TP2}[TP]{Trivial Pursuit}
\acro{TPCC}{Trivial Pursuit with Clipping Constraints}
\acro{TPE}{tree Parzen estimator}
\acro{TPR}{True Positive Rate}
\acro{TSE}{Taylor Series Expansion}
\acro{TS-VAD}{target speaker \acs{VAD}}
\acro{TVAR}{Time-varying Autoregression}
\acro{UAV}{Unmanned Aerial Vehicle}
\acro{UBM}{universal background model}
\acro{UDP}{User Datagram Protocol}
\acro{UEM}{unpartitioned evaluation map}
\acro{UFRJ}{Federal University of Rio de Janeiro}
\acro{UHF}{Ultra High Frequency}
\acro{UIS-RNN}{unbounded interleaved-state - recurrent neural network}
\acro{UKF}{Unscented Kalman Filter}
\acro{ULA}{Uniform Linear Array}
\acro{ULF}{Ultra Low Frequency}
\acro{UMTS}{Universal Mobile Telecommunications Service}
\acro{UPGMA}{unweighted pair group method with arithmetic mean}
\acro{US}{United States}
\acro{UTBM}{Universal Target Binary Mask}
\acro{VAD}{voice activity detection}
\acro{VB}{variational Bayes}
\acro{VBHMM}{variational Bayes hidden Markov model}
\acro{VBR}{Variable Bit-Rate}
\acro{VCV}{Vowel-Consonant-Vowel}
\acro{VRD}{Variance of Decay-rates}
\acro{VGC}{Voice Grade Channel}
\acro{vMF}{von Mises-Fisher}
\acro{VoIP}{voice over internet protocol}
\acro{VRT}{Vlaamse Radio- en Televisieomroeporganisatie\acroextra{. (Flemish Radio and Television Broadcasting Organization)}}
\acro{VSELP}{Vector Sum-excited Linear Prediction}
\acro{VST}{Virtual Studio Technology\acroextra{. An interface standard developed by Steinberg for adding plugins to an audio editor.}}
\acro{WADA}{Waveform Amplitude Distribution Analysis}
\acro{WASPAA}{Workshop on Applications of Signal Processing to Audio and Acoustics}
\acro{WAV}{Waveform Audio File Format}
\acro{WAVE}{Waveform Audio File Format}
\acro{WB}{wideband}
\acro{WER}{Word Error Rate}
\acro{WGN}{White Gaussian Noise}
\acro{WLAN}{Wireless \acs{LAN}}
\acro{WLLN}{Weak Law of Large Numbers}
\acro{WM}{Working Memory}
\acro{WMA}{Windows Media Audio}
\acro{WNG}{White Noise Gain}
\acro{WOLA}{weighted overlap-add}
\acro{WPE}{weighted prediction error}
\acro{WSS}{Weighted Spectral Slope}
\acro{WSOBM}{Weighted \acs{SOBM}}
\acro{WSTOI}{Weighted \acs{STOI}}
\acro{XAI}{explainable \acs{AI}}
\acro{XML}{explainable machine learning}
\acro{ZOS}{Zero-Order Statistics}
\end{acronym}

\maketitle

\begin{abstract}
This paper studies modulation spectrum features ($\mathbf{\Phi}$) and mel-frequency cepstral coefficients ($\mathbf{\Psi}$) in joint speaker diarization and identification (JSID).  JSID is important as speaker diarization on its own to distinguish speakers is insufficient for many applications, it is often necessary to identify speakers as well.  Machine learning models are set up using convolutional neural networks (CNNs) on $\mathbf{\Phi}$ and recurrent neural networks – long short-term memory (LSTMs) on $\mathbf{\Psi}$, then concatenating into fully connected layers.  

Experiment~1 shows machine learning models on both $\mathbf{\Phi}$ and $\mathbf{\Psi}$ have significantly better diarization error rates (DERs) than models on either alone; a CNN on $\mathbf{\Phi}$ has DER 29.09\%, compared to 27.78\% for a LSTM on $\mathbf{\Psi}$ and 19.44\% for a model on both.  Experiment~1 also investigates aleatoric uncertainties and shows the model on both $\mathbf{\Phi}$ and $\mathbf{\Psi}$ has mean entropy 0.927~bits (out of 4~bits) for correct predictions compared to 1.896~bits for incorrect predictions which, along with entropy histogram shapes, shows the model helpfully indicates where it is uncertain.

Experiment~2 investigates epistemic uncertainties as well as aleatoric using Monte Carlo dropout (MCD).  It compares models on both $\mathbf{\Phi}$ and $\mathbf{\Psi}$ with models trained on x-vectors ($\bm{\mathcal{X}}$), before applying Kalman filter smoothing on the epistemic uncertainties for resegmentation and model ensembles.  While the two models on $\bm{\mathcal{X}}$ perform better (DERs 10.23\% and 9.74\%) than the models on $\mathbf{\Phi}$ and $\mathbf{\Psi}$ (DER 17.85\%) after their individual Kalman filter smoothing, combining the models using a Kalman filter smoothing method improves the DER to 9.29\%.  Aleatoric uncertainties are again shown to be higher for incorrect predictions.

Both Experiments show models on $\mathbf{\Phi}$ do not distinguish overlapping speakers as well as anticipated.  However, Experiment~2 shows model ensembles do better with overlapping speakers than individual models do.

\end{abstract}

\begin{IEEEkeywords}
modulation spectrum, speaker diarization, aleatoric and epistemic uncertainty, Kalman filter smoothing.
\end{IEEEkeywords}

\IEEEpeerreviewmaketitle

\section{Introduction}
\IEEEPARstart{S}{peaker} diarization is the process of distinguishing different speakers in any given speech signal and identifying the times during which they speak. It involves two fundamental aspects: (i)~segmentation of speech data into either constant time periods (e.g. a fixed number of frames) or non-constant time periods that are homogeneous in some way (e.g. single speaker speech, overlapping speaker speech or no speech); and (ii)~clustering and/or labelling the segments identified to attribute them to individual speakers \cite{Park2022, AngueraMiro2012, Tranter2006, Soldi2016}.  Diarization is useful both alone to distinguish speakers and as an upstream process leading to speaker identification, \ac{ASR} or other systems.

Most diarization research systems distinguish speakers but do not identify them (e.g. as ``Speaker\_1'').  This is consistent with diarization challenges (e.g. the DIHARD challenges \cite{Ryant2018, Ryant2019, Ryant2020}), which usually also require the system to be tested on speakers not in the training set.  It is possible to have a subsequent speaker identification system that identifies speakers once they have been diarized, but a significant amount of research combines the two to, for example, improve performance or facilitate online applications.  Examples include speech separation using speaker inventory \cite{Wang2019, Han2020}, joint speaker identification and speech separation \cite{Park2022} and continuous speaker identification \cite{Flemotomos2020}.  Others reformulate as multi-label classification \cite{Fujita2020}. 

The term \ac{JSID} is used as the emphasis is on speaker diarization (including diarization scoring methods), but it also performs closed set speaker identification for training and testing features.

\subsection{Modulation Spectrum Background}\label{ss:modspecbg}

The modulation spectrum describes how the frequency content of a speech signal changes over time \cite{Hermansky2010}.  There are many different ways of calculating modulation spectrum features ($\mathbf{\Phi}$).  This paper uses the joint acoustic and modulation spectrum approach of \cite{Atlas2003} as extended in \cite{McKnight2021a}.  $\mathbf{\Phi}$ has two parts here: (i)~the \ac{ENV}, which uses \ac{AM} principles to look at the slowly changing temporal trajectory of specific acoustic frequency bands; and (ii)~the \ac{TFS}, which uses \ac{FM} principles to look at the rapidly changing instantaneous frequency around the centre frequencies of those acoustic frequency bands \cite{McKnight2021a}.  $\mathbf{\Phi}$ is expected to perform well for distinguishing speakers \cite{Zeng2005} and detecting overlapping speakers \cite{Atlas2003}.

\subsection{Uncertainty Quantification Background}

Most trained machine learning models are deterministic, with point values set for specific weights and outputs despite the inherent uncertainties caused by inaccurate labels and imperfect models.  Machine learning models that additionally indicate the confidence in their predictions are generally more advantageous.  Attempts have been made to quantify the levels of uncertainty \cite{Smith2014, Senge2014,  Acquesta2019, Jospin2021} in terms of: (a)~aleatoric uncertainty, also known as stochastic, statistical or irreducible uncertainty; and (b)~epistemic uncertainty, also known as systemic or reducible uncertainty.  Total uncertainty is both aleatoric and epistemic uncertainty.

Aleatoric uncertainty describes noise in observations, whereas epistemic uncertainty describes errors due to imperfect models.  Epistemic uncertainty describes uncertainty in the models fitted on the data, both for the weights of the models (i.e. parametric) and the model structures.  A \ac{BNN} is a common starting point for evaluating the parametric epistemic uncertainty by fitting probability distributions for some or all of the weights in the model \cite{Neal1996}, but it was found to be very difficult to find a suitable \ac{BNN} model architecture that did not underfit the data.  Consequently, this paper uses the Monte Carlo dropout approximation to Bayesian inference that applies dropout during testing as well as during training \cite{Gal2016a, Gal2016}.  Note that what one machine learning model treats as aleatoric uncertainty, another may treat as epistemic uncertainty, and \textit{vice versa} \cite{Kiureghian2009, Hullermeier2021}.  Research has only recently started to investigate uncertainty quantification in speaker diarization \cite{Silnova2020, Aronowitz2020}, and the research in those papers is very different from this research.

Aleatoric uncertainty is known to be particularly important for speaker diarization, and consequently \ac{JSID}, because of the difficulty and subjectivity in obtaining accurate labels \cite{McKnight2020a}.

\subsection{Resegmentation}

Most diarization systems use some form of resegmentation in the post-processing stage to improve initial clusters or labels \cite{Sell2015}.  This paper considers two: simple smoothing; and Kalman filter methods, both forward only and the two-pass \ac{RTS} fixed interval smoothing \cite{Brown2012}.

\subsection{This Research}

The novel contributions of this research are showing that: (a)~$\mathbf{\Phi}$ and \ac{MFCCs} both have information that the other does not have and establishing a model to take advantage of it; (b)~aleatoric uncertainties contain meaningful information that can be used to calculate frame entropies that show how models perform based on various conditions (e.g. the number of speakers in a frame); and (c)~total uncertainties can be used to combine models and improve results.

\section{Analysis}\label{s:modspecanalysis}

\subsection{Generating Modulation Spectrum Features $\mathbf{\Phi}$}

The 4-stage process used to generate $\mathbf{\Phi}$ is described in \cite{McKnight2021a}.  Here, they are: (a)~first \ac{STFT}; (b)~calculate the spectral envelope using Hilbert transforms; (c)~second \ac{STFT} across frequency bands; and (d)~calculate \ac{ENV} and \ac{TFS} features using Hilbert transforms.

As \cite{McKnight2021a} showed that the \ac{TFS} features contain additional speaker-specific information over and above the \ac{ENV} features, the form of \ac{TFS} features $\mathbf{\Phi}_{TFS} \in \mathbb{R}^{L \times K \times H}$ was chosen to have the same dimensions as the \ac{ENV} features $\mathbf{\Phi}_{ENV} \in \mathbb{R}^{L \times K \times H}$ where $L$ is the number of modulation frames, $K$ is the number of acoustic frequency bands and $H$ is the number of modulation frequency bands.  

$\mathbf{\Phi} \in \mathbb{R}^{L \times K \times H \times 2}$ stacks the \ac{ENV} and \ac{TFS} features so

\vspace{-1.2em}
\begin{align}
    \mathbf{\Phi}(l, k, h, 0) &= \mathbf{\Phi}_{ENV}(l, k, h) \text{ and }\\ \mathbf{\Phi}(l, k, h, 1) &= \mathbf{\Phi}_{TFS}(l, k, h)
\end{align}

\noindent for $\{l \in \mathbb{Z} : 0 \leq l \leq L-1\}$, $\{k \in \mathbb{Z} : 0 \leq k \leq K-1\}$ and $\{h \in \mathbb{Z} : 0 \leq h \leq H-1\}$.

To emphasise relative magnitudes within each frame rather than absolute values, each element $\mathbf{\Phi}_{ENV}(l, k, h)$ was divided by the Frobenius norm $||\mathbf{\Phi}_{ENV}(l)||_F$ of frame $l$, and similarly for $\mathbf{\Phi}_{TFS}(l)$.  As is common practice, the values were then standardised across frames.

\subsection{Uncertainty Quantification}\label{ss:uncqual}

\begin{figure*}[hbp!]
\vspace{-0.5em}
\centering
  \includegraphics[width=\linewidth]{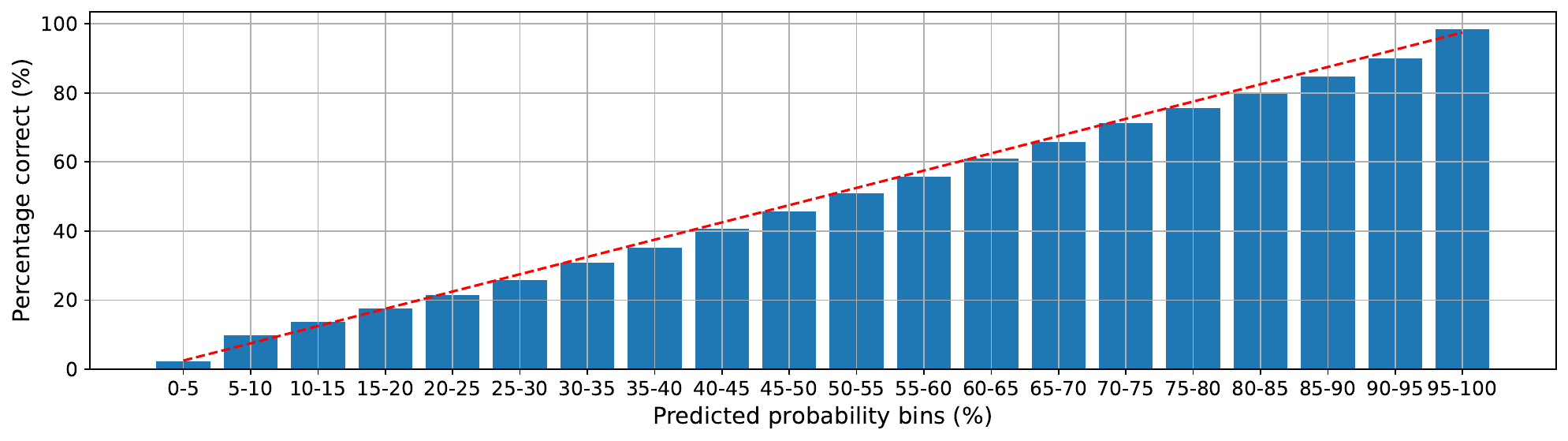}
\caption{Probability calibration graph on validation set for total uncertainties of MCD2-$\bm{\mathcal{X}}_R$ (defined in Section~\ref{ss:feats_systems}).}
\label{fig:calib}
\end{figure*}

The aleatoric uncertainty for a particular modulation frame $l$, speaker $s$ of $S$ possible speakers, input features $\mathbf{X}_{l, s}$ and output vector $\mathbf{y}_{l, s}$ where each speaker is a binary output is given by $p(\mathbf{y}_{l, s} | \mathbf{X}_{l, s}, \bm{\theta})$, where $\bm\theta$ denotes the weights of the relevant neural network comprising the weight terms $\mathbf{W}$ and bias terms $\mathbf{b}$.  The epistemic uncertainty is $p(\bm{\theta} | \mathcal{D}_{l, s})$, where $\mathcal{D}_{l, s}$ denotes the observed data, i.e. $\mathcal{D}_{l, s} = \{\mathbf{X}_{l, s}, \mathbf{y}_{l, s}\}$.

The experiments used Bernoulli output probability distributions following \cite{Webster} to give the additional option of taking random samples from the outputs. 
 The \texttt{IndependentBernoulli} class from the TensorFlow Probability library \cite{Dillon2017, Davidson-Pilon2015} was used.  However, the results reported in this paper only use the mean output, which gives exactly the same result as using sigmoid activation in the final dense layer when training with binary cross-entropy loss.  In addition, it was noted that the calibration graphs such as that shown in Fig.~\ref{fig:calib} were found to be well calibrated, so no probability calibration was used.

The output aleatoric probability predicted by a particular model for a particular speaker and modulation frame is denoted by $\hat{p}(\mathbf{y}_{l, s} | \mathbf{X}_{l, s}, \bm{\theta})$ and shortened to $\hat{p}_{l, s}$.  The model prediction for the speakers for that frame is then

\begin{equation}
    \hat{y}_{l, s} = 
    \left\{ 
        \begin{array}{ll}
        1 \quad \text{ if } \hat{p}_{l, s} > \lambda \\
        0 \quad \text{ if } \hat{p}_{l, s} \leq \lambda
        \end{array}\right.,
    \label{eq:threshold}
\end{equation}

\noindent where $\lambda$ is the probability threshold that is usually set to 0.5 (see Section~\ref{ss:threshold} for experiments on this).

Quantifying epistemic uncertainty using Monte Carlo dropout involves running the trained model $N$ times on the test data.  As the dropout is also applied when testing, the output values differ each time, and it is then possible to fit probability distributions for the predictions of each modulation frame that would quantify the uncertainty.  In this paper, the mean probability, 2.5\% to 97.5\% probability range, mean prediction and modal prediction are calculated.  Denoting each sample $n$ with superscript $^{[n]}$, the mean probability is 

\begin{equation}
    \bar{p}_{l, s} = \frac{1}{N}\sum_{n=0}^{N-1} \hat{p}_{l, s}^{[n]},
\end{equation}

\noindent the 2.5\% to 97.5\% percentile probability range is calculated from those same values $\hat{p}_{l, s}^{[n]}$ sorted in ascending order and the mean prediction is

\begin{equation}
    \bar{y}_{l, s} = \frac{1}{N}\sum_{n=0}^{N-1} \hat{y}_{l, s}^{[n]}.
\end{equation}

Defining the set of samples for each modulation frame and speaker as $\mathbb{\hat{Y}}_{l, s}$, the modal prediction is defined as
\begin{equation}
    \tilde{y}_{l, s} = \argmax_{\hat{y}_{l, s}^{[n]}} | \{\hat{y}_{l, s}^{[n]} \in \mathbb{\hat{Y}}_{l, s}\} | \,\, \forall \, n,
\end{equation}

\noindent where $|\{\cdot\}|$ denotes the cardinality of the set.

Since the Kalman filter discussed in Section \ref{ss:reseg} uses standard deviations as well as means, initial experiments used the means $\bar{p}_{l, s}$ (which were generally very close to the medians) and divided the 2.5\% to 97.5\% percentile ranges by four to approximate the standard deviation.  However, although this crude measure was a useful starting point, and better than a typical Gaussian distribution because the values of $\bar{p}_{l, s}$ were all within the range $[0, 1]$, better results were obtained by fitting a truncated Gaussian distribution $\phi_{l, s}$ instead.  There is no closed-form way to calculate the values, so the distributions were fitted using the \texttt{scipy.optimize.fmin\_slsqp} sequential least squares programming function on the \texttt{scipy.stats.truncnorm} generated values following \cite{scipy}.

\subsection{Resegmentation}\label{ss:reseg}

Initially, a simple smoothing mechanism was used.  This had two effects: (a)~it bridged prediction gaps ($\hat{y}_{l, s} = 1$ for P-MODEL and $\tilde{y}_{l, s} = 1$ for MCD-MODEL) of up to $G$ frames, where $\{G \in \mathbb{Z} : 0 \leq G \leq 10\}$, between the same person speaking were treated as all speech; and (b)~it then flattened any remaining single frame spikes.  The results shown in Table~\ref{tab:resultsPart1} had $G = 3$.  Although $G$ was initially tested as a hyperparameter to be optimised on the validation set, the optimum value of $G$ on the validation set did not always correspond to the optimum value of $G$ on the test set, which led to using $G = 3$ as an empirically selected value.

An alternative method that takes advantage of the standard deviations for the epistemic uncertainty is to use Kalman filters \cite{Brown2012}.  Kalman filters work on both single models and multiple models by treating the predictions of each model as separate observations.  Several variations of Kalman filters exist, but this paper uses the two-pass \ac{RTS} fixed interval method described in Algorithm~\ref{algo:kalman}.  The Kalman filter equations are normally set out in matrix form, but since this paper is only tracking a single variable $x$ that is based on the observations $z = \bar{p}_{l, s}$ the simpler formulation is preferred for clarity.  Some experiments were also run without the backward pass step of RTS, but results were generally better with it.

The validation set was used to find optimal values of certain Kalman filter hyperparameters for subsequent use on the training set.  Not all Kalman filter variable parameters were found to generalise well to the training set (e.g. the threshold percentage discussed in Section~\ref{ss:threshold}), so those were set at empirically selected fixed values.  Transition factor $f$ for frame $l$ and speaker $s$ is

\begin{equation}
    f_{l, s} = 
    \left\{ 
        \begin{array}{ll}
        f_0 \quad \text{ if forward and } \bar{p}_{l, s} > \lambda \\
        f_1 \quad \text{ if forward and } \bar{p}_{l, s} \leq \lambda \\
        f_2 \quad \text{ if backward}
        \end{array}\right.,
\end{equation}

\noindent where $f_0$, $f_1$ and $f_2$ are scalars that are fitted on the validation set. Some experiments tested setting $f_0$ and $f_1$ to be the same value, but it was generally found that $f_1$ should be higher than $f_0$.  This makes sense because although $f_0$ bears similarities to the probability that a speaker continues speaking and $f_1$ to the probability that a speaker starts speaking if not speaking before, it is not exactly the same because it (a)~is a multiple of the previous value of $\hat{x}_{l-1, s | l-1, s}$ and (b)~could not be greater than one ($p_{l, s}$ rapidly exceeded computer memory if it did).

The process variance $q_{l, s}$ is treated similarly, though only applies in the forward stage so

\begin{equation}
    q_{l, s} = 
    \left\{ 
        \begin{array}{ll}
        q_0 \quad \text{ if } \bar{p}_{l, s} > \lambda \\
        q_1 \quad \text{ if } \bar{p}_{l, s} \leq \lambda
        \end{array}\right..
\end{equation}

The observation factor $\mathbf{h} \in \mathbb{R}^{U \times 1}$ for $U$ models, where each $h_u$ is fitted on the validation set.  The model observation variance $R_{l, s} \in \mathbb{R}^{U \times U}$ is different for each modulation frame and is a diagonal matrix based on the epistemic uncertainty variance calculated using $\phi_{l, s}$ for each model $u$.  The variables in Algorithm~\ref{algo:kalman} are: (a)~$\hat{x}_{l, s|l-1, s}$ is the initial \textit{a priori} state estimate and $\hat{x}_{l, s|l-1, s}$ is the updated \textit{a posteriori} state estimate; (b)~$p_{l, s | l-1, s}$ is the initial \textit{a priori} covariance estimate and ~$p_{l, s | l, s}$ is the updated \text{a posteriori} state estimate (i.e. it should not be confused with the use of $p_{l, s}$ for probability in Section~\ref{ss:uncqual}; and (c)~$\textbf{k}_{l, s} \in \mathbb{R}^{U \times 1}$ is the Kalman gain.

\begin{algorithm}[!b]
  \SetAlgoLined
  \textit{Forward prediction} \\
  \For{$0 \leq s \leq S-1$}{
    \For{$0 \leq l \leq L-1$}{
      \textit{Predict step} \\
      {$\hat{x}_{l, s|l-1, s} = f_{l, s}\hat{x}_{l-1, s|l-1, s}$\\
      $p_{l, s|l-1, s} = f_{l, s}^2p_{l-1, s|l-1, s} + q_{l, s}$} \\
      \textit{Update step} \\
      {
            $\mathbf{\tilde{y}_{l, s}} = \mathbf{z}_{l, s} - \mathbf{h} \hat{x}_{l, s|l-1, s}$\\
            $\mathbf{S}_{l, s} = \mathbf{h} p_{l, s|l-1, s} \mathbf{h}^T + \mathbf{R}_{l, s}$\\
            $\mathbf{k}_{l, s} = p_{l, s|l-1, s} \mathbf{h}^T \mathbf{S}_{l, s}^{-1}$\\
            $\hat{x}_{l, s|l, s} = \hat{x}_{l, s|l-1, s} + \mathbf{k}_{l, s}^T \mathbf{\tilde{y}}_{l, s}$\\
            $p_{l, s|l, s} = (1 - \mathbf{k}_{l, s}^T \mathbf{h}) p_{l, s|l-1, s}$\\
        }
    }
  }
  \textit{Backward sweep} \\
    \For{$0 \geq s \geq S-1$}{
      \For{$L-2 \geq l \geq 0$}{
        $a_{l, s} = p_{l, s|l, s}f_{l, s}/p_{l+1, s|l, s}$ \\
        $\hat{x}_{l, s|L-1, s} = a_{l, s}(\hat{x}_{l+1, s|L-1, s} - \hat{x}_{l+1, s|l, s})$ \\
        $p_{l, s|L-1, s} = p_{l, s|l, s} + a_{l, s}(p_{l+1, s|L-1, s} - p_{l+1, s|l, s})/a_{l, s}$ \\
    }
  }
  \caption{Kalman filter tracking employed}
  \label{algo:kalman}
\end{algorithm}

\begin{table*}[htp!]
\begin{center}
\caption{Statistics for ES2008 meetings, where ``Dur.'' is duration of meeting, ``Tot. Spch'' is total speech time that includes overlapping speakers individually, ``Utts'' is number of utterances by individual speakers, ``Comb. Spch'' is speech time that only counts overlapping speakers once, ``Segs'' is number of speech segments (one segment may have more than one speaker and more than one utterance), ``Overlap'' is percentage of time that there is more than one speaker, ``Ch. Rate'' is speaker change rate (twice number of ground truth segments divided by speech file length) and ``ASD'' is average segment duration.}
\label{tab:ES2008stats}
\resizebox{\textwidth}{!}{
\begin{tabular}{|c|c|c|c|c|c|c|c|c|c|}
  \hline
    \multicolumn{1}{|c}{\textbf{Meeting}} &
    \multicolumn{1}{|c}{\textbf{Dur. (mins)}} &
    \multicolumn{1}{|c}{\textbf{Dur. (s)}} &
    \multicolumn{1}{|c}{\textbf{Tot. Spch (s)}} &
    \multicolumn{1}{|c}{\textbf{Utts}} &
    \multicolumn{1}{|c}{\textbf{Comb. Spch (s)}} &
    \multicolumn{1}{|c}{\textbf{Segs}} &
    \multicolumn{1}{|c}{\textbf{Overlap (\%)}} &
    \multicolumn{1}{|c}{\textbf{Ch. Rate (Hz)}} &
    \multicolumn{1}{|c|}{\textbf{ASD (s)}} \\
  \hline
    ES2008a & 17:23.360 & 1,043.360 &  806.640 &  168 & 775.940 & 107 & 3.955 & 0.322 & 4.80 \\
    ES2008b & 37:11.659 & 2,231.659 &  1,849.770 &  439 & 1,742.020 & 248 & 6.185 & 0.393 & 4.21 \\
    ES2008c & 35:02.621 & 2,102.621 & 1,957.110 & 396 & 1,741.850 & 174 & 12.358 & 0.377 & 4.94 \\
    ES2008d & 43:45.824 & 2,625.824 & 2,349.910 & 757 & 2,113.810 & 396 & 11.169 & 0.577 & 3.10 \\
  \hline
\end{tabular}
}
\end{center}
\end{table*}

\subsection{Scoring Metrics}\label{ss:scoring}

The standard binary accuracy metric in TensorFlow~2 was used on the training and validation sets, then the time-based \ac{DER} metric \texttt{md-eval.pl} developed for the \ac{NIST} Rich Transcription Challenges \cite{NIST2009} was used on the test set.  However, using the binary accuracy metric meant that a system would show around 78\% accuracy simply by predicting no speakers for any frames, so the precision and recall metrics give more meaningful results.

In addition to those metrics, it became evident when running the experiments that a frame-based \ac{DER} metric was essential.  Although it was possible to reconstruct the signal and apply \texttt{md-eval.pl} following each training epoch, it was prohibitively computationally intensive and slow.  By contrast, a frame-based \ac{DER} metric could take advantage of the vectorisation and parallelisation inherent in TensorFlow~2 and its batch training, and allow easier tracking of the \ac{DER} metrics during training of the models.  A further benefit of a frame-based \ac{DER} metric was to enable more detailed inspection of where errors occurred (e.g. whether the model tended to over- or under-estimate the number of speakers for a ground truth labelling of two or more speakers).  This Section \ref{ss:scoring} describes the construction of these frame-based \ac{DER} metrics.

The time-based \ac{DER} metric in \texttt{md-eval.pl} is

\vspace{-1em}
\begin{equation}
    DER_{\tau} = \frac{\tau_{M} + \tau_{FA} + \tau_{SE}}{\tau_{TOTAL}} = M_{\tau} + FA_{\tau} + SE_{\tau},
\end{equation}

\noindent where $\tau$ denotes time-based metrics (distinguished from $\epsilon$ used for the frame-based metrics later); $\{\tau_{M}, \tau_{FA}, \tau_{SE}, \tau_{TOTAL}\}$ are the missed speaker time, false alarm time, speaker error time and total speech time respectively; and $\{M_{\tau}, FA_{\tau}, SE_{\tau}\}$ are the time-based missed speaker percentage, false alarm percentage and speaker error percentage respectively.

For a particular batch, the ground truth training labels $\mathbf{Y}~\in~\mathbb{R}^{B~\times~S}$, where $B$ is the batch size, $S$ is the number of speakers and each entry is either 0 or 1 (this is not the same as one-hot encoding as there can be more than one speaker per modulation frame).  For each training batch, the miss error metric $\epsilon_{M}$ is calculated using

\begin{equation}
    \epsilon_{M} = \frac{1}{B} \sum\limits_{l=0}^{B-1} \max \Bigg(\sum\limits_{s=0}^{S-1}y_{l, s} - \sum\limits_{s=0}^{S-1}\hat{y}_{l, s},0 \Bigg)
\end{equation}

\noindent and the false alarm metric $\epsilon_{FA}$ is calculated similarly as

\vspace{-1em}
\begin{equation}
    \epsilon_{FA} =  \frac{1}{B} \sum\limits_{l=0}^{B-1} \max \Bigg(\sum\limits_{s=0}^{S-1}\hat{y}_{l, s} - \sum\limits_{s=0}^{S-1}y_{l, s},0 \Bigg).
\end{equation}

The Hadamard product $\odot$ is used to calculate $\mathbf{H}~=~\mathbf{Y}~\odot~\hat{\mathbf{Y}}$.  The speaker error metric $\epsilon_{SE}$ is

\vspace{-1em}
\begin{equation}
    \epsilon_{SE} = \frac{1}{B} \sum\limits_{l=0}^{B-1} \Bigg{[\min\Bigg({\sum\limits_{s=0}^{S-1}\hat{y}_{l, s}, \sum\limits_{s=0}^{S-1}y_{l, s}}\Bigg) - \sum\limits_{s=0}^{S-1}h_{l, s}\Bigg]}.
\end{equation}

These individual errors are summed to give $\epsilon_{DER}$ as the frame-based proxy for $\tau_{DER}$

\vspace{-1em}
\begin{equation}
    DER_{\epsilon} = \frac{\epsilon_{M} + \epsilon_{FA} + \epsilon_{SE}}{\epsilon_{TOTAL}} = M_{\epsilon} + FA_{\epsilon} + SE_{\epsilon},
\end{equation}

\noindent where $\epsilon$ denotes frame-based metrics; $\{\epsilon_{M}, \epsilon_{FA}, \epsilon_{SE}, \epsilon_{TOTAL}\}$ are number of missed speaker modulation frames, false alarm modulation frames, speaker error modulation frames and total speech modulation frames respectively; and $\{M_{\epsilon}, FA_{\epsilon}, SE_{\epsilon}\}$ are the frame-based missed speaker percentage, false alarm percentage and speaker error percentage respectively.  The frame-based values do not exactly match the time-based measures because of the rounding into modulation frames as well as the use of the \ac{GT-SAD} post-processing before calculating the time-based metrics.

\section{Experimental Design and Results}\label{s:experiments}

\subsection{Experiment Structure}

This research is split into two complementary experiments:

\begin{enumerate}
    \item \textbf{\textit{Experiment~1}} - studies whether modulation spectrum has useful information for speaker diarization beyond \ac{MFCCs} or $\mathbf{\Psi}$; and
    \item \textbf{\textit{Experiment~2}} - studies whether total uncertainty quantification can be used to gain more meaningful information about the model predictions and produce better models in the resegmentation step.
\end{enumerate}

\subsection{Datasets and Ground Truth Labels}\label{ss:datasets}

AMI Corpus \cite{Carletta2006} ES2008 headset recordings were used.  These four meetings all had the same four speakers, which enabled (a) for Experiment~1, training on three of them and testing on one and (b) for Experiment~2, training on two of them, validating and fitting resegmentation models on one and then testing on one.  Experiment~1 models were trained on meetings ES2008b, c and d, then tested on ES2008a as it is the shortest.  Experiment~2 models use the ES2008c meetings as the validation set as it is the second shortest.

Initially, ground truth labels were constructed from the \texttt{ES2008a.[A-D].segments.xml} files, but it was found that (a)~these contained silence of 0.25-0.5~s at the start and end of each segment \cite{Moore2005} which greatly increased miss errors as well as confusing the model training (the reason for these silences was possibly because false positives were seen as worse than false negatives in a \ac{SAD} used for \ac{ASR} \cite{Neo2022, Sohn1999}) and (b)~they included non-lexical sounds such as laughter and coughing.  Consequently, the ground truth labels, and also the ground truth \ac{SAD} which was used in Experiment~2, were constructed from the \texttt{ES2008a.[A-D].words.xml} files generated for the AMI corpus for words only using forced alignment and HTK \cite{Carletta2006} (conveniently already extracted in the ``only\_words'' directory of \cite{Landini2020, Landini2020c}).  This led to $DER_{\epsilon}$ and $DER_{\tau}$ improvements of around 6-8\%.  Table~\ref{tab:ES2008stats} shows general statistics for the meetings.  Diarization performance is greatly affected by the amount of overlapping speech as well as how often the speakers change.

\subsection{Features and Systems Used}\label{ss:feats_systems}

A particular challenge for comparing single frame vector features such as \ac{MFCCs} with modulation spectrum matrix features is the drastically different frame durations as well as the different shapes.  \ac{MFCCs} involve \ac{STFT}s that need stationary processes in each frame, so the frames cannot be longer than about 40~ms \cite{Paliwal2010a}.  By contrast, modulation spectrum features need many acoustic frames, so even if the acoustic frame steps $F_a$ used by the modulation spectrum are significantly shorter than those for \ac{MFCCs} denoted $F_{a2}$, using many of them will generally result in modulation frame steps $F_a$ being significantly longer than $F_{a2}$.  DiarTk \cite{Vijayasenan2012} is an existing speaker diarization system that is designed to work with multiple different input feature types, but unfortunately it (a)~requires different input features to have the same frame duration and step and (b)~only works on vector inputs.

\begin{table}[b]
\vspace{-0.6cm}
\begin{center}
\caption{P-MODEL structure (P-$\mathbf{\Phi}\mathbf{\Psi}$).}
\label{tab:cnnlstmstructure_alea}
\begin{tabular}{|c|c|c|c|c|}
  \hline
    & \textbf{Layer} & \textbf{Units/Filter} & \textbf{Activation} & \textbf{Output Shape}\\
  \hline
    \parbox[t]{2mm}{\multirow{7}{*}{\rotatebox[origin=c]{90}{Block 1}}} & Inputs & - & - & $(L, 25, 501, 2)$ \\
    & Conv2D & $F_1, (3, 3)$ & ReLU & $(L, 23, 499, F_1)$ \\
    & Conv2D & $F_2, (3, 3)$ & ReLU & $(L, 21, 497, F_2)$ \\
    & MaxPool2D & $(3, 3)$ & - & $(L, 7, 165, F_2)$ \\
    & Conv2D & $F_3, (3, 3)$ & ReLU & $(L, 5, 163, F_3)$ \\
    & Conv2D & $F_4, (3, 3)$ & ReLU & $(L, 3, 161, F_4)$ \\
    & MaxPool2D & $(3, 3)$ & - & $(L, 1, 53, F_4)$ \\
    & Flatten & - & - & $(L, 53 \times F_4)$ \\
  \hline
    \parbox[t]{2mm}{\multirow{3}{*}{\rotatebox[origin=c]{90}{Block 2}}} & Inputs & - & - & $(L, 25, 19)$ \\
    & BiLSTM & $U_L$ & tanh/sig. & $(L, 2U_{L})$ \\
    & Flatten & - & - & $(L, 2U_{L})$ \\
  \hline
    \parbox[t]{2mm}{\multirow{8}{*}{\rotatebox[origin=c]{90}{Combined}}} & Concat. & - & - & $(L, 53 \times F_4 + 2U_{L})$ \\
    & Dropout & - & - & - \\
    & Dense & $U_{D1}$ & ReLU & $(L, U_{D1})$ \\
    & Dropout & - & - & - \\
    & Dense & $U_{D2}$ & ReLU & $(L, U_{D2})$ \\
    & Dropout & - & - & - \\
    & Dense & $S$ & None & $(L, S)$ \\
    & Bernoulli & - & - & $(L, S)$ \\
  \hline
\end{tabular}
\end{center}
\end{table}

Consequently, a novel deep learning architecture extending \cite{McKnight2021a} was established that uses \ac{CNNs} for the modulation spectrum features and a many-to-one \ac{LSTM} for a number of \ac{MFCCs} such that the aggregate duration of those frame steps equals the modulation frame step.  The outputs of those \ac{CNN} and \ac{LSTM} blocks were then concatenated and fed into fully connected layers.  This architecture had a further benefit that it continued to work with just the \ac{CNN} or \ac{LSTM} blocks alone feeding into the fully connected layers.  Bidirectional \ac{LSTM} (BiLSTM) with output concatenation were tried in both Experiments~1 and 2, and were found to improve results in Experiment~1 but not Experiment~2.

For Experiment~1, $\mathbf{\Phi}$ was constructed from $W_a = 3~\text{ms}$ frames stepped by $F_a = 1~\text{ms}$ and modulation frames $W_m = 1~\text{s}$ stepped by $F_m = 250~\text{ms}$.  The \ac{MFCCs} $\mathbf{\Psi}$ were created using 30~ms frames stepped by 10~ms using 32 filter banks, reduced to 19 dimensions after the \ac{DCT} then grouped into lots of $\frac{250}{10} = 25$ so that the aggregate length of those frame steps equals $F_m$.

For Experiment~2, $F_m$ was increased to 1.5~s for consistency with the x-vector lengths in \cite{Landini2020a, Landini2020} and which are used in Experiment~2.  This enabled more direct comparison as well as facilitating the use of Kalman filters to combine the model outputs.  This paper denotes x-vectors generally as $\bm{\mathcal{X}}$.  There are two types of x-vectors used in this paper, namely the $512 \times 1$ dimension x-vectors from \cite{Landini2020a} defined as $\bm{\mathcal{X}_B}$ and the $256 \times 1$ dimension x-vectors from \cite{Landini2020a} defined as $\bm{\mathcal{X}_R}$.

\begin{table}[b]
\begin{center}
\caption{MCD-MODEL structure (MCD1-$\mathbf{\Phi}\mathbf{\Psi}$ and MCD2-$\mathbf{\Phi}\mathbf{\Psi}$); ``Act.'' is activation, ``Out. Sh.'' is output shape.}
\label{tab:cnnlstmstructure_alea_epiMC}
\resizebox{\linewidth}{!}{
\begin{tabular}{|c|c|c|c|c|c|}
  \hline
    & \textbf{Layer} & \textbf{Units/Filter} & \textbf{Act.} & \textbf{MCD1 Out. Sh.} & \textbf{MCD2 Out. Sh.}\\
  \hline
    \parbox[t]{2mm}{\multirow{11}{*}{\rotatebox[origin=c]{90}{Block 1}}} & Inputs& - & - & $(L, 25, 501, 2)$ & $(L, 25, 751, 2)$ \\
    & Conv2D & $F_1, (3, 3)$ & ReLU & $(L, 23, 499, F_1)$ & $(L, 23, 749, F_1)$ \\
    & Dropout & - & - & - & - \\
    & Conv2D & $F_2, (3, 3)$ & ReLU & $(L, 21, 497, F_2)$ & $(L, 21, 747, F_2)$ \\
    & Dropout & - & - & - & - \\
    & MaxPool2D & $(3, 3)$ & - & $(L, 7, 165, F_2)$ & $(L, 7, 249, F_2)$\\
    & Conv2D & $F_3, (3, 3)$ & ReLU & $(L, 5, 163, F_3)$ & $(L, 5, 247, F_3)$\\
    & Dropout & - & - & - & - \\
    & Conv2D & $F_4, (3, 3)$ & ReLU & $(L, 3, 161, F_4)$ & $(L, 3, 245, F_4)$\\
    & Dropout & - & - & - & \\
    & MaxPool2D & $(3, 3)$ & - & $(L, 1, 53, F_4)$ & $(1, 81, F_4)$\\
    & Flatten & - & - & $(L, 53 \times F_4)$ & $(L, 81 \times F_4)$\\
  \hline
    \parbox[t]{2mm}{\multirow{3}{*}{\rotatebox[origin=c]{90}{Block 2}}} & Inputs & - & - & $(L, 25, 19)$ & $(L, 25, 19)$\\
    & (Bi)LSTM & $U_L$ & tanh/sig. & $(L, 2U_L)$ & $(L, U_L)$ \\
    & Flatten & - & - & $(L, 2U_L)$ & $(L, U_L)$ \\
  \hline
    \parbox[t]{2mm}{\multirow{7}{*}{\rotatebox[origin=c]{90}{Combined}}} & Concat. & - & - & $(L, 53F_4 + 2U_L)$ & $(L, 81F_4 + U_L)$ \\
    & Dense & $U_{D1}$ & ReLU & $(L, U_{D1})$ & $(L, U_{D1})$ \\
    & Dropout & - & - & - & - \\
    & Dense & $U_{D2}$ & ReLU & $(L, U_{D2})$ & $(L, U_{D2})$ \\
    & Dropout & - & - & - & - \\
    & Dense & $S$ & None & $(L, S)$ & $(L, S)$ \\
    & Bernoulli & - & - & $(L, S)$ & $(L, S)$ \\
  \hline
\end{tabular}}
\end{center}
\end{table}

The models were constructed in Python using TensorFlow~2.  The TensorFlow Probability (TFP) library was used for the probabilistic layers \cite{Davidson-Pilon2015}.  The aleatoric aspects follow \cite{Webster} and Monte Carlo dropout follows \cite{Durr2020}.  Experiments on model architecture tuning showed most consistent and reliable performance for: (a)~the probabilistic model (P-MODEL) in Table~\ref{tab:cnnlstmstructure_alea}; and (b)~the Monte Carlo dropout models on \ac{MFCCs} and modulation spectrum features in Table~\ref{tab:cnnlstmstructure_alea_epiMC} and on x-vectors in Table~\ref{tab:xvecstructure_alea_epiMC} (generically referred to as the MCD-MODEL).  Using P-MODEL for both modulation spectrum features is shortened to P-$\mathbf{\Phi\Psi}$, for modulation spectrum features only (i.e. without Block~2) is shortened to P-$\mathbf{\Phi}$ and using it without Block~1 is shortened to P-$\mathbf{\Psi}$ for MFCCs only, P-$\mathbf{\Psi\Delta}$ with delta features as well and P-$\mathbf{\Psi\Delta\Delta}$ with both delta and delta-delta features as well.  Similarly, the MCD-MODEL used in Experiment~1 is shortened to MCD1-$\mathbf{\Phi\Psi}$, shortened to MCD1-$\mathbf{\Phi}$ without Block~2 and to MCD1-$\mathbf{\Psi}$, MCD1-$\mathbf{\Psi\Delta}$ and MCD1-$\mathbf{\Psi\Delta}$ without Block~1 depending on the delta features included.  The same logic applies to MCD2-$\mathbf{\Phi\Psi}$, MCD2-$\mathbf{\Phi}$, MCD2-$\mathbf{\Psi}$, MCD2-$\mathbf{\Psi\Delta}$ and MCD2-$\mathbf{\Psi\Delta\Delta}$.

\begin{figure*}[htp!]
  %\vspace{-0.5cm}
  \centering
  \includegraphics[clip, trim=0cm 0cm 0cm 0cm, width=\textwidth]{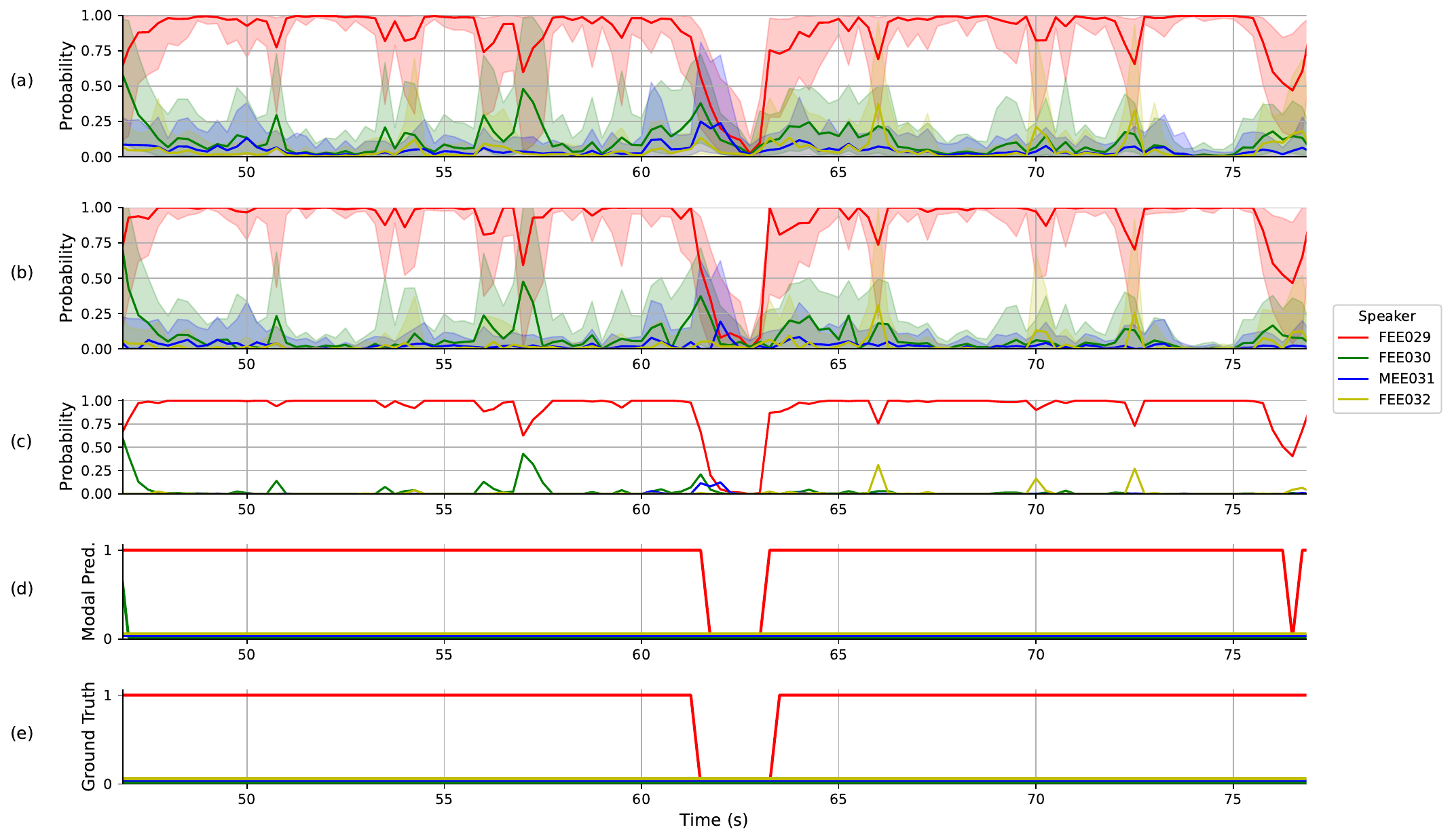}
  \caption{30~s extract of ES2008a for MCD1-$\mathbf{\Phi\Psi}$ before resegmentation: (a)~aleatoric uncertainty from the mean $\bar{p}_{l, s}$ and epistemic uncertainty 2.5\% to 97.5\% percentile range; (b)~total uncertainty from fitting truncated Gaussian $\phi_{l, s}$; (c)~mean prediction $\bar{y}_{l, s}$; (d)~modal prediction $\tilde{y}_{l, s}$; and (e)~ground truth. (d) and (e) are offset slightly on y-axis to clarify overlaps.  The shaded regions in (a) and (b) show the epistemic uncertainty ranges).}
  \label{fig:aleaepidrop}
  \vspace{-0.3cm}
\end{figure*}

The input data was put into batches of 512 shuffled each epoch, ``valid'' padding applied to the \ac{CNNs}, the \ac{LSTM} not stateful across modulation frames and adaptive momentum (adam) optimisation used.  $N = 200$ for MCD-MODELs.

Fig.~\ref{fig:aleaepidrop} shows MCD1-$\mathbf{\Phi\Psi}$ example outputs for a 30~s test extract.  The epistemic uncertainty ranges are clearly visible in parts (a) and (b), then the mean and modal predictions in (c) and (d) respectively look similar to the ground truth in (e).

Monte Carlo dropout has additional dropout layers after each CNN layer.  Following \cite{Gal2016}, the \ac{LSTM} has the same dropout value for the input as well as for the recurrent layers (not shown in Table~\ref{tab:xvecstructure_alea_epiMC} as included in LSTM function).  Importantly, dropout continued to be applied in validation and testing.  In each case, the model was trained using up to 100 epochs (patience 10 and early stopping 25) and up to 100 \ac{TPE} trials.

\begin{table}[b]
%\vspace{-0.6cm}
\begin{center}
\caption{MCD2-$\bm{\mathcal{X}}_B$ and MCD2-$\bm{\mathcal{X}}_R$.}
\label{tab:xvecstructure_alea_epiMC}
\begin{tabular}{|c|c|c|c|}
  \hline
    \textbf{Layer} & \textbf{Filter/Units} & \textbf{Activation} & \textbf{Output Shape}\\
  \hline
    Inputs & - & - & $(L, [512|256], 1)$ \\
    Dense & $U_{D1}$ & ReLU & $(L, U_{D1})$ \\
    Dropout & - & - & - \\
    Dense & $U_{D2}$ & ReLU & $(L, U_{D2})$ \\
    Dropout & - & - & - \\
    Dense & $U_{D3}$ & None & $(L, S)$ \\
    Bernoulli & - & - & $(L, S)$ \\
  \hline
\end{tabular}
\end{center}
\end{table}

Experiment~2 compares results to three baseline systems: (a)~DiarTk on $\mathbf{\Psi}$ \cite{Vijayasenan2012}, which is a agglomerative information bottleneck system; (b)~the x-vector based system in \cite{Landini2020a} (BDII); and (c)~the x-vector based system in \cite{Landini2020} (ResNet101).  

\subsection{Generating Features}

Because the features are computed with different frame sizes, the relevant speech signals were prepended and appended with enough zeros before extracting the relevant features to ensure alignment.  Accordingly, for a speech signal of time duration $T$, the number of modulation (and x-vector) frames was $N_m~=~ \lceil\frac{T}{F_m}\rceil$.  Speech signals were prepended with $[(W_m-F_m)~+~(W_a-F_a)]f_s/2$ zeros and appended with $[(N_m \times F_m - T) + (W_m-F_m)~+~(W_a-F_a)]f_s/2$ zeros before calculating $\mathbf{\Phi}$ and $\bm{\mathcal{X}}$.  A copy of original speech signal were prepended with $(W_{a2}-F_{a2})/2$ zeros and appended with $[(N_m \times F_m - T) + (W_{a2}-F_{a2})]f_s/2$ zeros before calculating $\mathbf{\Psi}$.

Although desirable to avoid using any \ac{SAD}, uncertainties around when the speech actually started meant that there were long lead-in periods that could be mitigated using a post-processing \ac{SAD}, applied to the ground truth data.  Furthermore, the x-vectors were trained on speech only, and therefore not trained to distinguish speech from non-speech.  The frame-based error metrics are reported before this \ac{GT-SAD} is performed, so only the time-based error metrics are affected by it.  Note that there is a significant advantage in using a \ac{SAD} as pre-processing as the system can infer at least one speaker in the relevant segment, which biases results in favour of systems using them.

\subsection{Data Augmentation and Dither}

For Experiment~2, because the size of the training was comparatively small, data augmentation was used to double the size of the training set by supplementing the clean data with features generated from the same speech files but with \ac{AWGN}.  The noisy signal for each sample was generated using random Gaussian $\sim N(0, 1)$ and scaled to achieve the desired target signal-to-noise ratio (SNR), then added to the speech signal with values still in 16-bit \texttt{wav} format (i.e. before dividing by 32,768).  Using 30~dB SNR, which would be almost imperceptible when listening \cite{ICSI2000}, was found to improve test DER of the $\bm{\mathcal{X}}$-based models by 1-2\%, though was less helpful for the $\mathbf{\Phi}$-based models.  Using 20~dB SNR in initial experiments made results worse in all cases, which suggests some sensitivity of the methods to noise.

\begin{table*}[hbp!]
\vspace{-0.3cm}
\begin{center}
\caption{Experiment~1 test set results comparison on best models of each type (all in \%): (a)~``Accu.'' is accuracy, ``Prec.'' is precision and ``Rec.'' is recall; (b)~no resegmentation applied; and (c)~post-processing \ac{GT-SAD} applies to time-based errors.}
\label{tab:resultsPart1}
\resizebox{\textwidth}{!}{
\begin{tabular}{|c|c||c|c|c|c||c|c|c|c||c|c|c|c|}
  \hline
    & \multicolumn{1}{p{1.5cm}||}{~~~~~\textbf{Model}} & \multicolumn{1}{p{0.9cm}|}{~\textbf{Accu.}} & \multicolumn{1}{p{0.9cm}|}{~~\textbf{Prec.}} & \multicolumn{1}{p{0.8cm}|}{\hspace{5pt}\textbf{Rec.}\textbf{}} & \multicolumn{1}{p{0.8cm}||}{~~~\textbf{F1}} & \multicolumn{1}{p{0.7cm}|}{$~~\textbf{M}_{\boldsymbol\epsilon}$} & \multicolumn{1}{p{0.8cm}|}{$~~\textbf{FA}_{\boldsymbol\epsilon}$} & \multicolumn{1}{p{0.8cm}|}{$~~\textbf{SE}_{\boldsymbol\epsilon}$} & \multicolumn{1}{p{1.0cm}||}{$~~\textbf{DER}_{\boldsymbol\epsilon}$} & \multicolumn{1}{p{0.7cm}|}{$~~\textbf{M}_{\boldsymbol\tau}$} & \multicolumn{1}{p{0.8cm}|}{$~~\textbf{FA}_{\boldsymbol\tau}$} & \multicolumn{1}{p{0.8cm}|}{$~~\textbf{SE}_{\boldsymbol\tau}$} & \multicolumn{1}{p{1.0cm}|}{$~~\textbf{DER}_{\boldsymbol\tau}$}  \\
  \hline
    \parbox[t]{2mm}{\multirow{5}{*}{\rotatebox[origin=c]{90}{Aleatoric}}} & P-$\mathbf{\Psi}$ & 93.26 & 90.87 & 72.21 & 80.47 & 18.04 & 2.25 & 3.33 & 23.62 & 23.75 & 0.03 & 4.00 & 27.78 \\
    & P-$\mathbf{\Psi} \Delta$ & 93.42 & 91.71 & 72.34 & 80.88 & 18.52 & 2.28 & 2.76 & 23.55 & 24.33 & 0.00 & 3.25 & 27.58 \\
    & P-$\mathbf{\Psi} \Delta\Delta$ & 93.29 & 92.07 & 71.21 & 80.31 & 19.31 & 1.89 & 2.83 & 24.03 & 25.38 & 0.00 & 3.38 & 28.76 \\
    & P-$\mathbf{\Phi}$ & 92.08 & 84.73 & 71.71 & 77.68 & 15.24 & 3.43 & 6.52 & 25.18 & 20.39 & 0.22 & 8.48 & 29.09 \\
    & P-$\mathbf{\Phi}\mathbf{\Psi}$ & 94.80 & 90.38 & 81.65 & 85.79 & 10.85 & 3.43 & 3.26 & 17.54 & 14.71 & 0.81 & 3.92 & 19.44 \\
    \cline{1-14}
    \parbox[t]{2mm}{\multirow{5}{*}{\rotatebox[origin=c]{90}{Total}}} & MCD1-$\mathbf{\Psi}$ & 93.47 & 90.46 & 73.83 & 81.30 & 17.06 & 2.92 & 3.07 & 23.05 & 23.15 & 0.24 & 3.49 & 26.88 \\
    & MCD1-$\mathbf{\Psi} \Delta$ & 93.81 & 92.43 & 73.83 & 82.09 & 17.73 & 2.25 & 2.40 & 22.38 & 23.39 & 0.28 & 2.79 & 26.47 \\
    & MCD1- $\mathbf{\Psi} \Delta\Delta$ & 93.67 & 91.39 & 74.05 & 81.81 & 17.13 & 2.54 & 2.83 & 22.50 & 22.60 & 0.35 & 3.26 & 26.21 \\
    & MCD1-$\mathbf{\Phi}$ & 92.57 & 87.08 & 72.02 & 78.84 & 16.22 & 2.92 & 5.29 & 24.44 & 21.50 & 0.32 & 6.93 & 28.75 \\
    & MCD1-$\mathbf{\Phi}\mathbf{\Psi}$ & 94.85 & 93.81 & 78.38 & 85.40 & 14.59 & 1.94 & 2.04 & 18.57 & 19.44 & 0.62 & 2.40 & 22.45 \\
  \hline
\end{tabular}
}
\end{center}
\end{table*}

Although dither was applied in the published code for generating $\bm{\mathcal{X}}$ \cite{landini2020b}, for this research dither was disapplied.  Additional experiments were also carried out on Experiment~2 $\bm{\mathcal{X}}$, $\mathbf{\Phi}$ and $\mathbf{\Psi}$ to add dither (specifically a random integer in the range $[-4:4]$ added to the 16-bit \texttt{wav} speech signal), but results were not affected significantly and are not reported here.

\subsection{Selecting Results}

The experiments in this paper used the Tree Parzen Estimator (TPE) in the Hyperopt library \cite{Bergstra2013, Hutter2011} to optimize the hyperparameters.  For Experiment~1, the model with lowest $DER_{\epsilon}$ after each \ac{TPE} trial was saved.  For Experiment~2, the model from each \ac{TPE} trial was then run 200 times in the Monte Carlo test simulation (applying dropout), and the modal prediction of each of the 200 modulation frames was used to calculate the final model prediction for that \ac{TPE} trial.  The best model retained from all \ac{TPE} trials was the one that had the lowest $DER_{\epsilon}$ on the validation set after the Monte Carlo simulations, so models that could have had lower $DER_{\epsilon}$ on the test set were discarded (avoids possible overfitting).

The search space comprised the number of CNN filters in each layer ($F_1, F_2, F_3, F_4 \in \{16, 32, 64, 128\}$), the number of units in each LSTM layer ($U_L \in \{128, 256, 512\}$) and each dense layer ($U_{D1}, U_{D2} \in \{32, 64, 128, 256\}$), the dropout rates (uniform in $[0, 1]$) and the learning rate (log uniform in $[10^{-5}, 10^{-3}]$).  With the LSTMs, the input dropout and the recurrent dropout were specified to be the same value following \cite{Gal2016}.  Others that were tried were regularization rates, though later removed to rely on dropout only.  It was found that the model optimization was non-convex and the random initialisation of the weights and biases had a corresponding significant impact on results.  To mitigate this, the best hyperparameters for the number of CNN filters, the number of units in each LSTM layer and dense layer and the dropout rates were identified before running again with those hyperparameters fixed.

\begin{figure*}[!hb]
  \vspace{-0.5cm}
  \centering
  \includegraphics[clip, trim=0cm 0cm 8.5cm 0cm, width=0.8\textwidth]{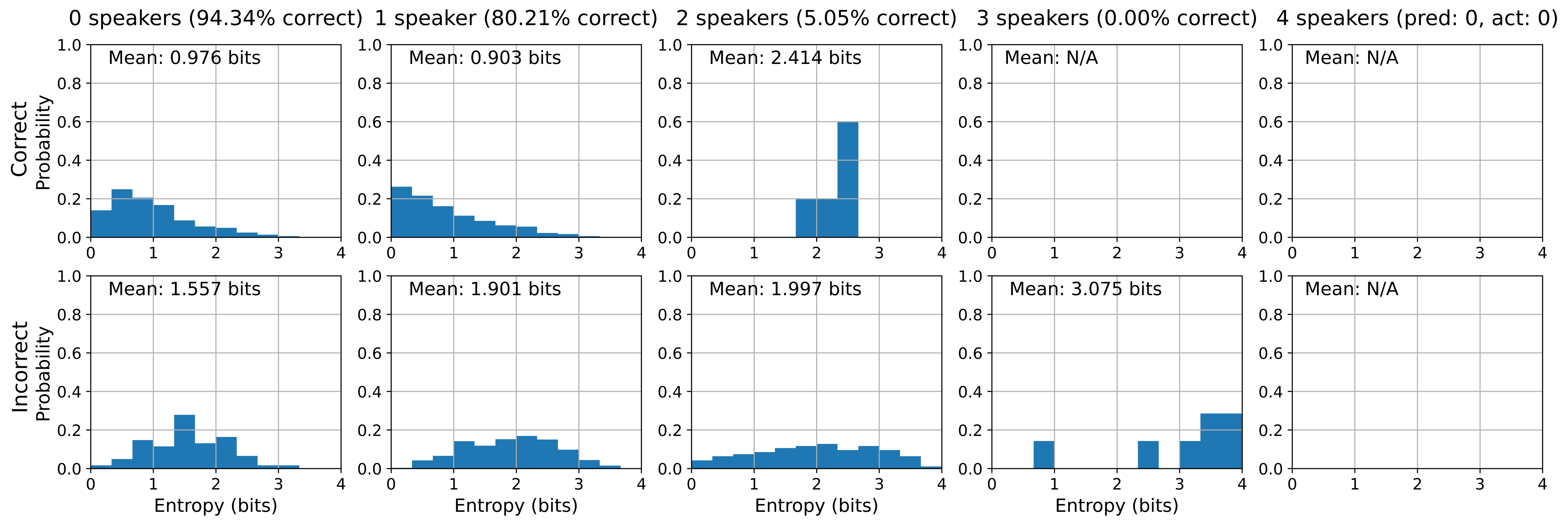}
  \caption{Experiment~1 MCD1-$\mathbf{\Phi\Psi}$ entropies histograms for correct and incorrect modulation frame predictions broken up by actual number of speakers in those modulation frames (none had 4 speakers).}
  \label{fig:entropiesNumSpkrs}
\end{figure*}

\begin{figure}[!t]
  \centering
  \includegraphics[width=\linewidth]{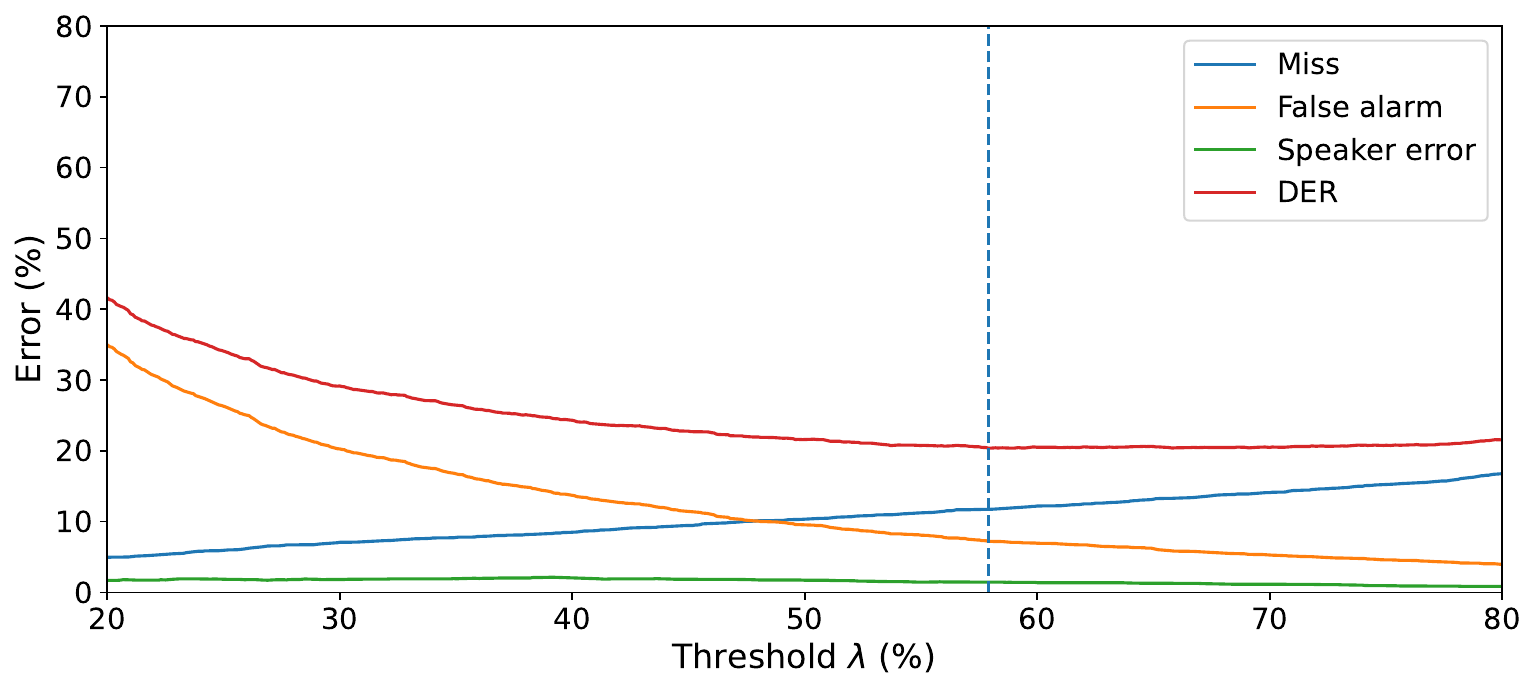}
  \caption{Frame-based errors for threshold in 0.1\% increments.}
  \label{fig:thresholds}
  \vspace{-0.3cm}
\end{figure}

\subsection{Threshold}\label{ss:threshold}

Plotting the output frame-based error graphs against the prediction threshold $\lambda$ in $0.1\%$ increments consistently showed behaviour such as seen in Fig.~\ref{fig:thresholds}, with the best test \ac{DER} ranging from 35\% to 65\%.  It was found that the optimum threshold for the validation set was generally not a good predictor of the optimum threshold for the test set, so it was not advantageous to include the threshold as a hyperparameter to be optimised in the \ac{TPE}.  Instead, $\lambda$ was set at 0.5.

\subsection{Experiment~1 Results: $\mathbf{\Psi}$ v $\mathbf{\Phi}$}\label{ss:MFCCsvPhi}

Table~\ref{tab:resultsPart1} shows how the best performing model performed on the relevant features on which it was trained (see the ``Model'' columns).  All models had better precision than recall, and misses were by far the largest error type.  $M_{\epsilon}$ was generally around 60-80\% of $DER_{\epsilon}$ and $M_{\tau}$ was generally around 70-90\% of $DER_{\tau}$ (the latter is greater than the former because using the \ac{GT-SAD} effectively reduces $FA_{\tau}$ to near zero each time, thereby increasing overall $DER_{\tau}$ attributable to $M_{\tau}$).  This shows that false negatives are more prevalent than false positives and misses are more problematic than false alarms.

The most significant result is that models using both $\mathbf{\Psi}$ and $\mathbf{\Phi}$ together gives dramatically better results than either alone.  This is particularly significant for the aleatoric models where P-$\mathbf{\Phi\Psi}$ has $DER_{\epsilon}$ 17.54\% and $DER_{\tau}$ 19.44\% is a substantial improvement of both P-$\mathbf{\Psi}$ $DER_{\epsilon}$ 23.62\% and $DER_{\tau}$ 27.78\% and P-$\mathbf{\Phi}$ $DER_{\epsilon}$ 25.18\% and $DER_{\tau}$ 29.09\%.  The total uncertainty models also show significant improvements using both $\mathbf{\Psi}$ and $\mathbf{\Phi}$ features, though not quite as dramatic as MCD1-$\mathbf{\Phi\Psi}$ that has $DER_{\epsilon}$ 18.57\% and $DER_{\tau}$ 22.45\% improving from MCD1-$\mathbf{\Psi}$ $DER_{\epsilon}$ 23.05\% and $DER_{\tau}$ 26.88\% and MCD1-$\mathbf{\Phi}$ $DER_{\epsilon}$ 24.44\% and $DER_{\tau}$ 28.75\%.  There are two possible conclusions to draw from this: (a)~that modulation spectrum features have additional information about speaker identity that is not present in \ac{MFCCs} and \textit{vice versa}; and/or (b) the method of extracting the information using either \ac{CNN} or \ac{LSTM} extracts the information in a different way that effectively provides additional information about speaker identity; the \ac{CNN} takes advantage of information in adjacent modulation spectrum features in through the $3 \times 3$ filters whereas the \ac{LSTM} uses sequential information in the \ac{MFCCs}.

\begin{figure}[!t]
  \centering
  \includegraphics[width=\linewidth]{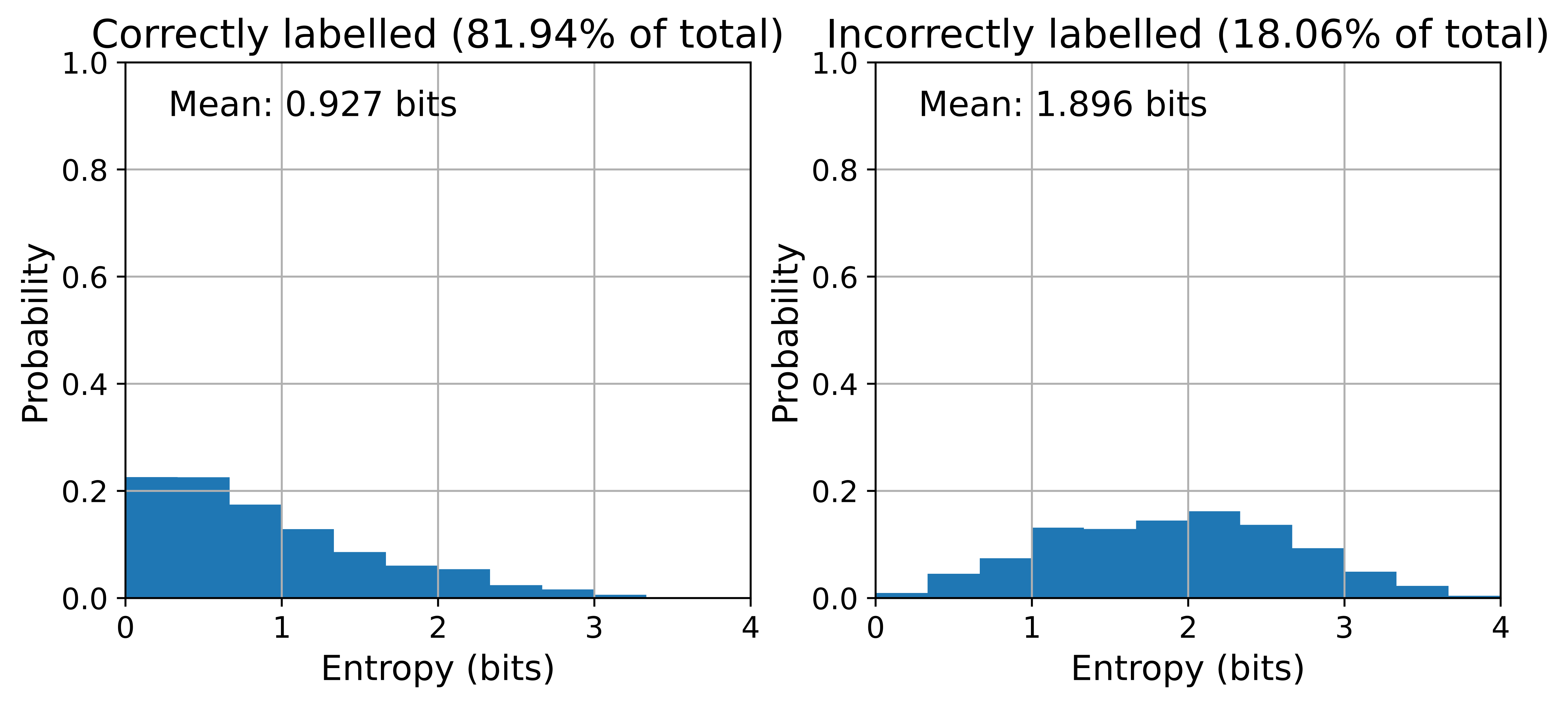}
  \caption{Experiment~1 MCD1-$\mathbf{\Phi\Psi}$ entropies histograms for correct and incorrect predictions of modulation frames.}
  \label{fig:entropiesCorrectIncorrect}
  \vspace{-0.3cm}
\end{figure}

Other Table~\ref{tab:resultsPart1} results of interest are: (a)~using \ac{MFCCs} on their own gives slightly better performance than $\mathbf{\Phi}$ alone; (b)~adding delta coefficients ($\Delta$) improves results marginally, but adding delta-delta coefficients ($\Delta\Delta$) is less reliable as it sometimes improves results and sometimes does not; and (c)~total uncertainty models on either $\mathbf{\Phi}$ or $\mathbf{\Psi}$ alone give slightly better results than the aleatoric only models on those features, but models on both $\mathbf{\Phi}$ and $\mathbf{\Psi}$ show the opposite.

\begin{figure*}[hbp!]
  \vspace{-0.5cm}
  \centering
  \includegraphics[clip, trim=0cm 0cm 0cm 0cm, width=\textwidth]{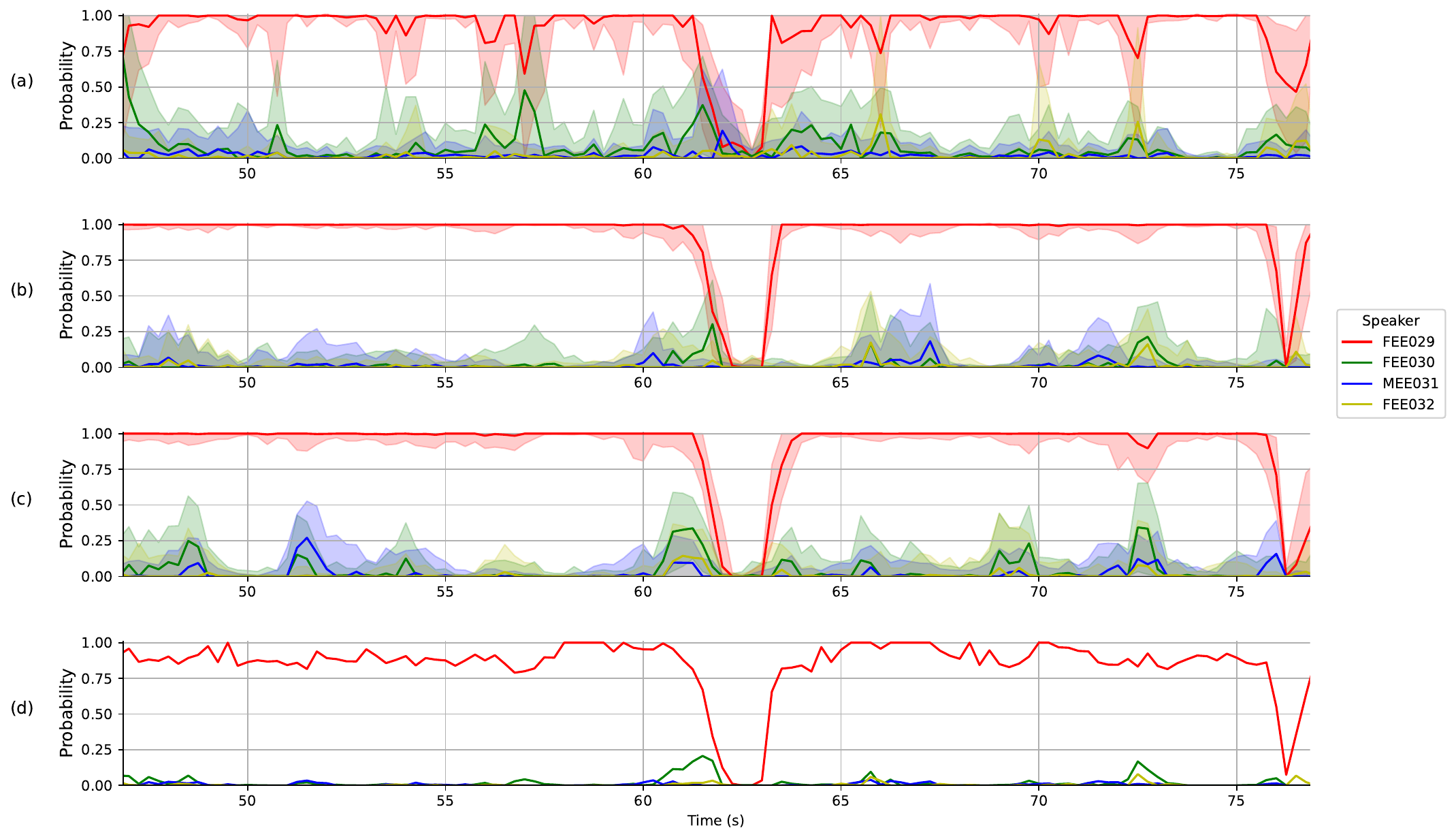}
  \caption{Experiment~2 30~s extract of test meeting ES2008a for total uncertainty of fitting the truncated Gaussian distributions for: (a)~MCD2-$\mathbf{\Phi\Psi}$; (b)~MCD2-$\bm{\mathcal{X}_B}$; and (c)~MCD2-$\bm{\mathcal{X}_R}$.  (d)~is the Kalman filter combined and smoothed version MCD2-$\mathbf{\Phi\Psi}, \bm{\mathcal{X}_B}, \bm{\mathcal{X}_R}$, which only has a single aleatoric uncertainty prediction per modulation frame.}
  \label{fig:aleaepidrop2}
\end{figure*}

Entropies for right and wrong predictions in Fig.~\ref{fig:entropiesCorrectIncorrect} clearly show higher entropies for the wrong predictions as the mean is 0.927~bits for the correct predictions compared to 1.896~bits for the incorrect predictions.  It is also evident that the shape of the correct predictions histogram is right-skewed whereas the incorrect predictions histogram is bell-shaped, which shows that in many cases the model rightly indicates that it is uncertain.  There are four speakers in these files, so the maximum entropy for each modulation frame is four.

Distinguishing entropies based on the number of actual speakers in each modulation frame as shown in Fig.~\ref{fig:entropiesNumSpkrs} suggests that the modulation spectrum is not as good at distinguishing overlapping speakers as might be anticipated.  For the modulation frames with 0 or 1 speaker, the means and histogram shapes for the correct predictions are similar to the correct predictions from Fig.~\ref{fig:entropiesCorrectIncorrect}, and similarly those for the incorrect predictions are similar to the incorrect predictions from Fig.~\ref{fig:entropiesCorrectIncorrect}.  This is not the case with 2 speakers as the mean entropy for the correct predictions is higher than for the incorrect predictions, and many of the incorrect predictions show a high confidence (i.e. low entropy) in their results despite being incorrect.  None of the modulation frames with 3 speakers were correctly predicted, and the mean entropy was high thereby correctly indicating the uncertainty.  Equivalent graphs for other best performing models were found to be similar so not shown.

\subsection{Experiment~2 Results: $\bm{\mathcal{X}}$ v $\mathbf{\Phi}$}\label{ss:xVecvPhi}

\begin{figure*}[htp!]
  \centering
  \includegraphics[clip, trim=0cm 19.8cm 0cm 0cm, width=\textwidth]{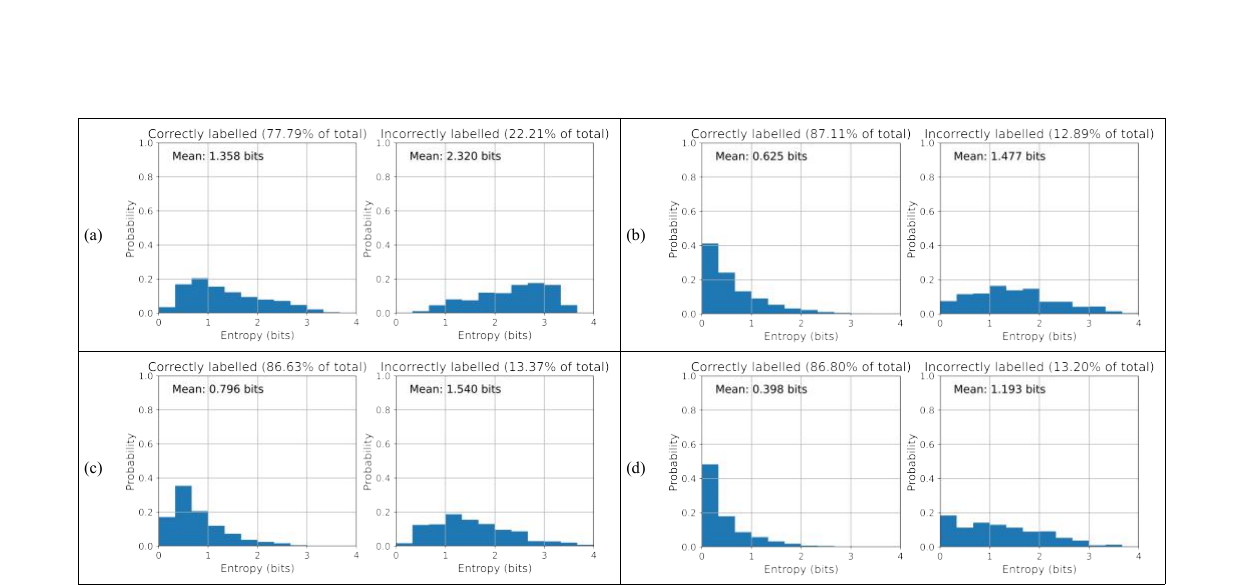}
  \caption{Experiment~2 entropies histograms for correct and incorrect predictions of modulation frames for: (a)~MCD2-$\mathbf{\Phi\Psi}$; (b)~MCD2-$\bm{\mathcal{X}_B}$; (c)~MCD2-$\bm{\mathcal{X}_R}$; and (d) the Kalman filter combined and smoothed version MCD2-$\mathbf{\Phi\Psi}, \bm{\mathcal{X}_B}, \bm{\mathcal{X}_R}$.}
  \label{fig:entropies}
  \vspace{-0.3cm}
\end{figure*}

The ``No Reseg.'' section of Table~\ref{tab:resultsPart2} shows how the best performing model performed on the relevant data on which it was trained as specified in the ``Model'' column before any resegmentation was applied.  The first three models in that section (MCD2-$\mathbf{\Phi\Psi}$, MCD2-$\bm{\mathcal{X}_B}$ and MCD2-$\bm{\mathcal{X}_R}$) were created for this paper.  The last three are baseline models, but their respective resegmentation methods were disapplied.  Fig.~\ref{fig:aleaepidrop2} compares a 30~s extract of each of these models, showing that the ones based on $\bm{\mathcal{X}}$ have significantly greater confidence in their predictions and will consequently dominate the Kalman filter combination, and (d) shows the Kalman filter combination before taking the predictions based on the chosen threshold $\lambda$.  Fig.~\ref{fig:entropies} shows the entropies histograms for correct and incorrect predictions.

MCD2-$\mathbf{\Phi\Psi}$ performed worse than P-$\mathbf{\Phi\Psi}$ and MCD1-$\mathbf{\Phi\Psi}$ from Experiment~1.  Although using 1.5~s modulation frames might have been expected to give better results, here the problems were that (a)~there was less training data (meeting ES2008c was used as the validation set in Experiment~2) and data augmentation was not as good a substitute and (b)~the wider modulation frames meant more labelling uncertainty.

Unsurprisingly, the two models based on $\bm{\mathcal{X}}$ did better than the one based on $\mathbf{\Phi}$ and $\mathbf{\Psi}$ as (a)~they have the benefit of much more training data for a wider range of speakers and (b)~the data is cleaner in that the speakers are clearly speaking for nearly all of the relevant frames (this is the way VoxCeleb 1 and 2 are set up, there is no timing information to consider).  

\begin{table*}[hbp!]
\vspace{-0.3cm}
\begin{center}
\caption{Experiment~2 epistemic test set results comparison on best performing models of each type (all in \%), where: (a)~``Accu.'' is accuracy, ``Prec.'' is precision, ``Rec.'' is recall; and (b) no frame-based scores available for the baseline systems.}
\label{tab:resultsPart2}
\resizebox{\textwidth}{!}{
\begin{tabular}{|c|c||c|c|c|c||c|c|c|c||c|c|c|c|}
  \hline
     & \multicolumn{1}{p{1.5cm}||}{~~~~~~~~~~~~~~~~\textbf{Model}} & \multicolumn{1}{p{0.9cm}|}{~\textbf{Accu.}} & \multicolumn{1}{p{0.9cm}|}{~~\textbf{Prec.}} & \multicolumn{1}{p{0.8cm}|}{\hspace{5pt}\textbf{Rec.}\textbf{}} & \multicolumn{1}{p{0.8cm}||}{~~~\textbf{F1}} & \multicolumn{1}{p{0.7cm}|}{$~~\textbf{M}_{\boldsymbol\epsilon}$} & \multicolumn{1}{p{0.8cm}|}{$~~\textbf{FA}_{\boldsymbol\epsilon}$} & \multicolumn{1}{p{0.8cm}|}{$~~\textbf{SE}_{\boldsymbol\epsilon}$} & \multicolumn{1}{p{1.0cm}||}{$~~\textbf{DER}_{\boldsymbol\epsilon}$} & \multicolumn{1}{p{0.7cm}|}{$~~\textbf{M}_{\boldsymbol\tau}$} & \multicolumn{1}{p{0.8cm}|}{$~~\textbf{FA}_{\boldsymbol\tau}$} & \multicolumn{1}{p{0.8cm}|}{$~~\textbf{SE}_{\boldsymbol\tau}$} & \multicolumn{1}{p{1.0cm}|}{$~~\textbf{DER}_{\boldsymbol\tau}$}  \\
  \hline
    \parbox[t]{2mm}{\multirow{6}{*}{\rotatebox[origin=c]{90}{No Reseg.}}} & MCD2-$\mathbf{\Phi} \mathbf{\Psi}$ & 93.60 & 90.69 & 76.17 & 82.80 & 16.10 & 3.14 & 3.19 & 22.42 & 20.69 & 0.19 & 2.91 & 23.79 \\
    & MCD2-$\bm{\mathcal{X}}_{B}$ & 96.44 & 94.13 & 87.89 & 90.91 & 9.06 & 3.69 & 0.74 & 13.49 & 9.30 & 0.66 & 0.88 & 10.84 \\
    & MCD2-$\bm{\mathcal{X}}_{R}$ & 96.34 & 93.55 & 87.98 & 90.68 & 9.03 & 4.22 & 0.69 & 13.94 & 9.29 & 0.33 & 0.73 & 10.34 \\
    & DiarTk $\mathbf{\Psi\Delta\Delta}$ & - & - & - & - & - & - & - & - & 3.80 & 0.00 & 11.39 & 15.33 \\
    & BDII & - & - & - & - & - & - & - & - & 3.84 & 0.00 & 5.68 & 9.52 \\
    & ResNet101 & - & - & - & - & - & - & - & - &  3.84 & 0.00 & 12.26 & 16.10 \\
  \hline
    \parbox[t]{2mm}{\multirow{3}{*}{\rotatebox[origin=c]{90}{Smoo.}}} & MCD2-$\mathbf{\Phi}\mathbf{\Psi}$ & 95.07 & 93.06 & 81.73 & 87.03 & 13.80 & 3.95 & 0.98 & 18.74 & 14.28 & 1.03 & 1.27 & 16.58 \\
    & MCD2-$\bm{\mathcal{X}}_{B}$ & 96.52 & 92.81 & 89.76 & 91.26 & 7.59 & 4.94 & 0.69 & 13.22 & 7.42 & 0.53 & 0.91 & 8.86 \\
    & MCD2-$\bm{\mathcal{X}}_{R}$ & 96.42 & 92.02 & 90.11 & 91.06 & 7.23 & 5.61 & 0.72 & 13.61 & 7.22 & 0.27 & 0.71 & 8.21\\
  \hline
    \parbox[t]{2mm}{\multirow{5}{*}{\rotatebox[origin=c]{90}{KF}}} & MCD2-$\mathbf{\Phi}\mathbf{\Psi}$ & 94.58 & 87.73 & 85.11 & 86.40 & 9.37 & 6.95 & 2.68 & 19.00 & 10.54 & 4.01 & 3.30 & 17.85 \\
    & MCD2-$\bm{\mathcal{X}}_{B}$ & 96.54 & 92.02 & 90.79 & 91.40 & 6.56 & 5.49 & 0.89 & 12.94 & 7.10 & 2.02 & 1.11 & 10.23 \\
    & MCD2-$\bm{\mathcal{X}}_{R}$ & 96.48 & 91.64 & 90.91 & 91.28 & 6.66 & 6.01 & 0.69 & 13.37 & 6.89 & 1.84 & 1.01 & 9.74 \\
    & MCD2-$\bm{\mathcal{X}}_B, \bm{\mathcal{X}}_{R}$ & 96.54 & 92.69 & 90.02 & 91.34 & 7.28 & 4.96 & 0.79 & 13.03 & 7.34 & 1.00 & 0.93 & 9.27 \\
    & MCD2-$\mathbf{\Phi}\mathbf{\Psi}, \bm{\mathcal{X}}_B, \bm{\mathcal{X}}_{R}$ & 96.39 & 91.61 & 90.47 & 91.03 & 6.92 & 5.92 & 0.79 & 13.63 & 6.77 & 1.59 & 0.93 & 9.29 \\
  \hline
    \parbox[t]{2mm}{\multirow{3}{*}{\rotatebox[origin=c]{90}{Base}}} & DiarTk $\mathbf{\Psi\Delta\Delta}$ Viterbi & - & - & - & - & - & - & - & - & 3.80 & 0.13 & 12.34 & 16.28 \\
    & BDII VBx & - & - & - & - & - & - & - & - & 3.84 & 0.00 & 7.41 & 11.25 \\
    & ResNet101 VBx & - & - & - & - & - & - & - & - & 3.84 & 0.00 & 2.39 & 6.23 \\
  \hline
\end{tabular}}
\end{center}
\end{table*}

As with Experiment~1, all models had better precision than recall and misses were by far the largest component of errors.

Table~\ref{tab:resultsPart1} compares results with the three baseline systems.  In each case, the results are reported both with and without their respective in-built resegmentation methods.  The baseline systems are all designed to undercluster before their resegmentation, though in the case of DiarTk and BDII the initial clustering gave good results that were subsequently made worse by the resegmentation (no change was made to the tuned hyperparameters of those systems).  Each of these systems are for unsupervised speaker diarization and have a significant advantage in that they use the ground truth \ac{SAD} before clustering the frames generated.  Nonetheless, in general the models tested in this paper perform better than DiarTk and BDII, though not quite as well as the more recent ResNet101.

\subsection{Experiment~2 Results: Resegmentation and Ensembles}\label{ss:reseg_combined}

The need for some form of resegmentation is evident when inspecting the graphical results (e.g. comparing the predictions in Fig.~\ref{fig:aleaepidrop}(d) with the ground truth in Fig.~\ref{fig:aleaepidrop}(e)) and noting the high proportion of the overall $DER_{\epsilon}$ and $DER_{\tau}$ attributable to $M_{\epsilon}$ and $M_{\tau}$ respectively in both Experiments.  

Table~\ref{tab:resultsPart2} in the ``Smoo.'' section shows how applying simple smoothing improves the results.  The improvement in the frame-based measures is significant and comparable to those for Kalman filter smoothing, but the particularly good performance comes for the time-based measures.  This dramatic improvement is largely attributable to the application of post-processing \ac{GT-SAD}.  Because the \ac{GT-SAD} reduces $FA_{\tau}$ significantly, it will also reduce $DER_{\tau}$ significantly unless $M_{\tau}$ and $SE_{\tau}$ increase significantly.  $M_{\tau}$ and $SE_{\tau}$ do often increase as there is reduced speech duration to be assessed and consequently they will increase unless they are specifically improved (this is clear from Table~\ref{tab:resultsPart1}).  By smoothing troughs of up to $G = 3$ and flattening spikes within the \ac{GT-SAD} (which by definition has at least one speaker), $M_{\tau}$ and $SE_{\tau}$ turn out to be similar to $M_{\tau}$ and $SE_{\tau}$ here, thereby resulting in $DER_{\tau}$ being significantly better than $DER_{\epsilon}$.  This simple smoothing results in better performance than the DiarTk and BDII methods, but somewhat lower than ResNet101.

The results in Table~\ref{tab:resultsPart2}, ``KF'', show how applying Kalman filter smoothing also improves results.  The frame-based metrics improve for all the single models, though most significantly for MCD2-$\mathbf{\Phi\Psi}$ (22.42\% to 19.00\%) as MCD2-$\bm{\mathcal{X}_B}$ (13.49\% to 12.94\%), and MCD2-$\bm{\mathcal{X}_R}$ (13.94\% to 13.37\%) only improved marginally.  Again, the improvement in time-based metrics was more substantial in all cases, with MCD2-$\mathbf{\Phi\Psi}$ 23.79\% to 17.85\%, MCD2-$\bm{\mathcal{X}_B}$ 10.84\% to 10.23\% and 10.34\% to 9.74\%, albeit not as good as for simple smoothing.

The combined models MCD2-$\bm{\mathcal{X}_B}, \bm{\mathcal{X}_R}$ and MCD2-$\mathbf{\Phi\Psi}, \bm{\mathcal{X}_B}, \bm{\mathcal{X}_R}$ do not improve the best $DER_{\epsilon}$, but do improve the $DER_{\tau}$.  This improvement largely comes from the reduced $FA_{\tau}$ after applying the \ac{GT-SAD}.  Using the Kalman filter smoothing to combine models results in decent performance, but not as good as simple smoothing or the best baseline model.  This is most likely because the models being combined are giving results that are too similar -- if one had great precision and one had great recall, then combining them would make more sense than combining two with better precision than recall.  A similar argument applies to misses, false alarms and speaker errors.  Disappointingly, MCD2-$\mathbf{\Phi\Psi}$ did not give the improved detection of overlapping speakers anticipated.

The entropies histograms in Fig.~\ref{fig:entropies} show that MCD2-$\bm{\mathcal{X}_B}$ has the most modulation frames correct at 87.11\%.  This is different from the metrics in Table~\ref{tab:resultsPart2} as it looks at the predictions of all speakers in the modulation frame, not summing each individually.  The shape of the MCD2-$\bm{\mathcal{X}_B}$ correct predictions histogram is strongly right skewed and has a low mean of 0.625~bits that is less than half the 1.477~bits mean of the incorrect predictions.  The incorrect predictions histogram is also slightly right skewed, which makes it difficult to distinguish correct and incorrect predictions based on the entropies alone.  The MCD2-$\bm{\mathcal{X}_R}$ histograms and figures show similar patterns and numbers to $\bm{\mathcal{X}_B}$, although with somewhat reduced performance.  The MCD2-$\mathbf{\Phi\Psi}$ correct predictions mean is 1.358~bits, significantly less than the 2.320~bits of the incorrect predictions, but higher than those of the other models.  However, the fact that the correct predictions histogram is right skewed whereas the incorrect predictions histogram is left skewed is a significant and advantageous difference.  The Kalman filter combined and smoothed model MCD2-$\mathbf{\Phi\Psi}, \bm{\mathcal{X}_B}, \bm{\mathcal{X}_R}$ has 86.80\% correct predictions, slightly lower than the 87.11\% of MCD2-$\bm{\mathcal{X}_B}$, but the mean entropy of the correct predictions is substantially lower at 0.398~bits compared to 0.625~bits and, importantly, is just 33.4\% of the mean entropy of the correct predictions compared to 42.3\% of MCD2-$\bm{\mathcal{X}_B}$ so correct predictions should be easier to distinguish from incorrect predictions.  The combined model's correct entropies histogram is more strongly right skewed than the others and its entropy information is more informative.

The entropies histograms equivalent to Fig.~\ref{fig:entropiesNumSpkrs} are not reproduced for Experiment~2 for conciseness, but as in Experiment~1, the results for 0 and 1 speaker are similar to the general findings seen in Fig.~\ref{fig:entropies}.  None of the models handled 2 speakers well.  The MCD2-$\mathbf{\Phi\Psi}$ had only 0.66\% correct compared to 3.29\% for MCD2-$\bm{\mathcal{X}_B}$ and 1.97\% for MCD2-$\bm{\mathcal{X}_R}$.  However, the Kalman filter combined and smoothed version improved this substantially to 10.53\% correct, suggesting that model ensembles should indeed help with overlapping speakers, and it also gave the highest percentage of correct frames with 1 speaker at 93.15\%, though the number of correct predictions for 0 speaker frames fell substantially to 80.08\%.  No models correctly predicted frames with 3 speakers.

Lastly, the baselines all had much better $M_{\tau}$ than the MCD2-MODELS, but the latter had much better $SE_{\tau}$.  This is likely because the baselines used the \ac{GT-SAD} as a pre-processing step and were therefore able to infer that there was at least one speaker in the relevant segments (the $M_{\tau}$ were all due to missed overlapping speakers, which the baseline systems were not able to detect).  By contrast, the MCD2-MODELS used \ac{GT-SAD} as a post-processing step, so if they predicted a speaker when the \ac{GT-SAD} said there was none then those would be removed, but predicting no speaker when the \ac{GT-SAD} said there was one did not help.

\section{Discussion and Conclusion}\label{s:conclusion}

Experiment~1 clearly shows that models using both $\mathbf{\Phi}$ and $\mathbf{\Psi}$ are better than models using either alone for both probabilistic and Monte Carlo dropout models: (a)~$DER_{\tau}$ is 29.09\% for P1-$\mathbf{\Phi}$, 27.78\% for P1-$\mathbf{\Psi}$ and 19.44\% for P1-$\mathbf{\Phi\Psi}$; and (b)~$DER_{\tau}$ is 28.75\% for MCD1-$\mathbf{\Phi}$, 26.88\% for MCD1-$\mathbf{\Psi}$ and 22.45\% for MCD1-$\mathbf{\Phi\Psi}$.  Experiment~1 also shows that the model on both features has mean entropy 0.927~bits (maximum 4~bits) for its correct predictions compared to 1.896~bits for its incorrect predictions, which along with the entropy histogram shapes shows the model helpfully indicates where it is uncertain.

Experiment~2 shows that models on $\bm{\mathcal{X}}$ ($DER_{\tau}$ is 10.23\% for MCD2-$\bm{\mathcal{X}_B}$ and 9.74\% for MCD2-$\bm{\mathcal{X}_R}$) perform better than models on both $\mathbf{\Phi}$ and $\mathbf{\Psi}$ ($DER_{\tau}$ 17.85\% for MCD2-$\mathbf{\Phi\Psi}$), in each case after their individual Kalman filter smoothing.  Combining the models using a Kalman filter smoothing method improves the $DER_{\tau}$ to 9.29\% for MCD2-$\mathbf{\Phi\Psi}, \bm{\mathcal{X}_B}, \bm{\mathcal{X}_R}$, which shows it to be an advantageous way of combining models, though performance still lags behind simple smoothing of individual models ($DER_{\tau}$ 8.21\% for MCD2-$\bm{\mathcal{X}_R}$).  The aleatoric and epistemic uncertainties are again shown to be higher for incorrect predictions.

Both Experiments~1 and 2 show that models on the modulation spectrum are not as good at distinguishing overlapping speakers as anticipated.  This could be because the relevant \ac{CNN} model structure is insufficient to identify the relevant relations between modulation spectrum features (only a simple $3 \times 3$ \ac{CNN} filter with 4 layers is used) rather than a failing with the modulation spectrum.  However, Experiment~2 shows that the combined model identifies overlapping speakers substantially better than the individual models does, though the accuracy is still poor at 10.53\%.

\bibliographystyle{IEEEtran}

\bibliography{sapstrings, bibl}

\end{document}